\documentclass[12pt]{iopart}

\usepackage{iopams}
\usepackage{graphics,epsfig}
\usepackage{datetime}
\expandafter\let\csname equation*\endcsname\relax
\expandafter\let\csname endequation*\endcsname\relax
\usepackage{amsmath}
\usepackage{color}

\newcommand{\bin}[2]{{#1 \choose #2}}
\newcommand{\nn}{\nonumber}
\newcommand{\ba}{\begin{equation}}
\newcommand{\ea}{\end{equation}}
\newcommand{\bbb}{\begin{align}}
\newcommand{\eee}{\end{align}}
\newcommand{\bb}{\begin{eqnarray}}
\newcommand{\ee}{\end{eqnarray}}

\renewcommand{\eref}[1]{Eq.~(\ref{#1})}
\newcommand{\efig}[1]{Fig.~\ref{#1}}

\newcommand{\ff}{\makebox{\small $\frac{1}{2}$}}
\newcommand{\fff}[2]{\makebox{\small $\frac{#1}{#2}$}}

\renewcommand{\d}{{\rm d}}

\newcommand{\Ai}{\textrm{Ai}}
\newcommand{\Bi}{\textrm{Bi}}
\newcommand{\yy}{-y}
\newcommand{\bv}{{\textbf v}}
\newcommand{\diag}{{\textrm{diag}}}
\newcommand{\coupling}{\epsilon_g}
\newcommand{\ad}{a^{\dag}}
\newcommand{\bd}{b^{\dag}}
\newcommand{\ds}{\delta_0}

\begin{document}

\title{Transmission and tunneling probability in two-band metals: influence of magnetic breakdown on the Onsager phase of
quantum oscillations}
\author{Jean-Yves Fortin$^1$, Alain Audouard$^2$}

\address{$^1$Institut Jean Lamour, D\'epartement de Physique de la Mati\`ere
et des Mat\'eriaux, Groupe de Physique Statistique, CNRS - Nancy-Universit\'e
BP 70239 F-54506 Vandoeuvre les Nancy Cedex, France
\\ \ead{jean-yves.fortin@univ-lorraine.fr}
$^2$Laboratoire National des Champs Magn\'{e}tiques
Intenses (UPR 3228 CNRS, INSA, UGA, UPS) 143 avenue de Rangueil,
F-31400 Toulouse, France
\\ \ead{alain.audouard@lncmi.cnrs.fr}}
\today

\begin{abstract}

Tunneling amplitude through magnetic breakdown (MB) gap is considered for two
bands Fermi surfaces illustrated in many organic metals. In particular, the
S-matrix associated to the wave-function transmission through the MB gap for
the relevant class of differential equations is the main object allowing the
determination of tunneling probabilities and phases. The calculated
transmission coefficients include a field-dependent Onsager phase. As a result,
quantum oscillations are not periodic in $1/B$ for finite magnetic breakdown
gap. Exact and approximate methods are proposed for
computing ratio amplitudes of the wave-function in interacting two-band models.

\end{abstract}

%
\section{\label{sec:intro}Introduction}
%
In recent years, interest regarding determination of the quantum oscillations
phase has been renewed. This was in particular motivated by the observation of a
Berry phase both in three-dimensional metals~\cite{Mi99} and topological
insulators~\cite{Fuchs2010}, for example in the case of Dirac
fermions~\cite{Wr13}. One might add the effect of non-parabolicity of the
dispersion equation which, both in conventional fermions and, especially, in
Dirac fermions is liable to induce phase offsets~\cite{Fortin2015}.

The problem of the Onsager phase was nevertheless addressed much earlier,
regarding the effect of the phase offset induced by magnetic breakdown (MB)
~\cite{Slutskin:1967,Slutskin:1968,Huang:1976}. The case of the model Fermi
surface (FS), known as the linear chain of coupled orbits by Pippard
\cite{Pi62}, is addressed in Refs.~\cite{Slutskin:1967,Slutskin:1968}. As it is
well known, the first experimental realization of this FS topology was observed
in the organic conductor $\kappa$-(ET)$_2$Cu(SCN)$_2$, where ET stands for the
bis-ethylenedithio-tetrathiafulvalene molecule~\cite{Os88}. In addition to the
$\pi/2$ dephasing occurring at each MB reflection, it was demonstrated that a
field-dependent phase offset should be observed~\cite{Slutskin:1967} as it has
been checked for $\theta$-(ET)$_4$CoBr$_4$(C$_6$H$_4$Cl$_2$)~\cite{Au13}.

%
\begin{figure*}
\centering
\includegraphics[width=0.6\textwidth, clip,angle=0]{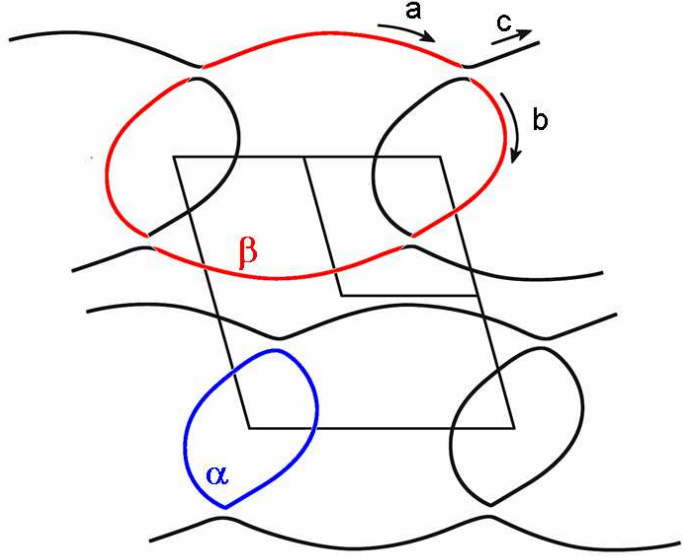}
\caption{
Fermi surface of organic conductor (BEDO-TTF)$_5$[CsHg(SCN)$_4$]$_2$
(from~\cite{Lyubovskii:2008}). An incoming wave (a) on the $\beta$-orbit is
reflected in (c) and transmitted to the $\alpha$-orbit (b).}
\label{fig0}
\end{figure*}
%

The main objective of this article is to consider the tunneling
phenomena in interacting cyclotronic orbits, and its implication to the
wavefunction characteristics at high and low field limits.
In the first step of this paper, we review the problem of transmission and
reflection coefficients within the S-matrix theory, when a particle coming
from infinity is scattered by a tunneling region. From the simple model due
to Rosen-Zener~\cite{Rosen:1932} and applied later to the magnetic breakdown
case~\cite{Slutskin:1967,Chambers:1968}, we focus on the effect of phase
divergence in the S-matrix amplitudes.
This actually occurs in different fields of physics, for example the
level-crossing problem~\cite{Torosov:2011}. Amplitude ratio of the wave
function is then considered in the second step when multiple paths are involved
in the tunneling process, leading to an oscillatory behavior of the
transmission coefficient. High field and semi-classical results are presented
and compared to the numerical resolution of the Schr\"odinger equation.
In the third step, we consider an exact approach to compute the quantum states
in the interacting case of two circular orbits with bound state conditions.
This new method is based on an extension of the usual (creation and
annihilation) bosonic operators of the harmonic
oscillator that includes effective coupling between the individual Fermi
surfaces using two parameters, representing the coupling itself and the gap
separately.
This is an approach that can be easily generalized to a linear chain of
coupled orbits, and which should give new insights on the wavefunction
properties. Finally, consequences on experimental de Haas-van Alphen
oscillations phase offset are considered for real FS of organic conductors.
%
\section{\label{sec:model}Review of the transmission phenomena in a simple
two-band model}

\begin{figure*}
\centering
\includegraphics[width=0.6\textwidth, clip,angle=0]{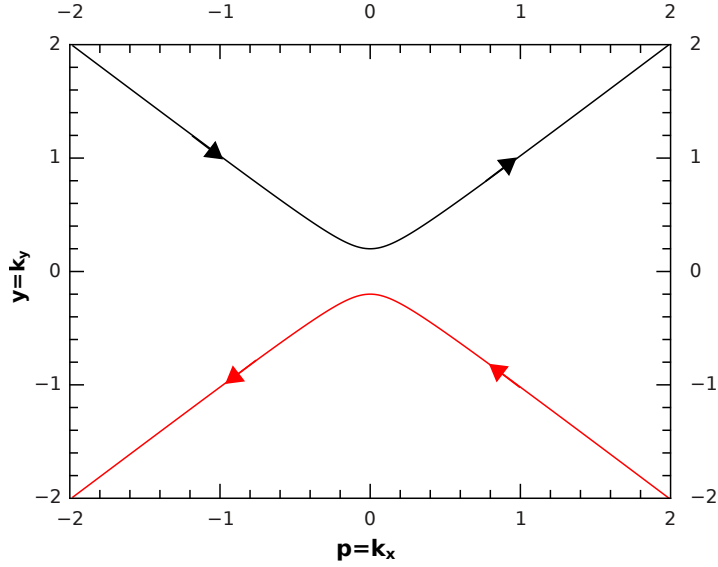}
\caption{
Effective two-band model. The hybridization parameter is $\coupling=0.2$.
The arrows represent the increase or decrease of the phase, specifically the
gradient of $\pm y^2/2h$. Here are represented two electronic bands
with trigonometric orientation of the trajectories.}
\label{fig1}
\end{figure*}

The presented model is intended to review the local transmission phenomena in
two-band metals with MB junctions, the FS of which achieves a linear chain of
coupled orbits (see
e.g.~\cite{Os88,Au12,Au15}). A typical example of such Fermi surface is
presented in~\efig{fig0} for
(BEDO-TTF)$_5$[CsHg(SCN)$_4$]$_2$~\cite{Lyubovskii:2008} (BEDO-TTF stands for the bis-ethylenedioxi-tetrathiafulvalene molecule), where an incoming
amplitude (a) is transmitted to (b) and reflected to (c). At the vicinity of the
MB junction, two linear
sheets hybridized with energy constant $\coupling$ can be
considered. The local Fermi surface is represented
on~\efig{fig1} for a non-zero coupling, and the linearized effective
Hamiltonian can be written as
\ba\label{model1}
\hat H
\left (
\begin{array}{c}
\varphi_1 \\ \varphi_2
\end{array}
\right )
=
\left (
\begin{array}{cc}
k_y+k_x & \coupling \\
\coupling & k_y-k_x
\end{array}
\right )
\left (
\begin{array}{c}
\varphi_1 \\ \varphi_2
\end{array}
\right )
=
\left (
\begin{array}{c}
0 \\ 0
\end{array}
\right )
\ea
For $\coupling=0$, the two sheets and the wavefunctions $\varphi_1$ and
$\varphi_2$
are independent. In such case, the MB gap which is proportional to
$\coupling^2$, is zero.
In presence of a magnetic field, the quantum representation of
this model is chosen such that $y=k_y$ and $\hat x=\hat k_x=2i\pi b\partial_y$,
with $b=eB/(2\pi \hbar)$.
In this case, the differential equations for the wavefunctions are
\bb
(y+ih\frac{\partial}{\partial y})\varphi_1+\coupling\varphi_2=0
\;\;\textrm{and}\;\;
\coupling\varphi_1+(y-ih\frac{\partial}{\partial y})\varphi_2=0
\ee
where $h=2\pi b$ is an effective magnetic Planck constant~\footnote{$h$ is not
to be confounded with the real Planck constant that we will write $2\pi\hbar$
in the rest of the paper}.  This set of first-order differential equations
can be reduced using the transformation $\varphi_1=\e^{iy^2/2h}g_1(y)$ and
$\varphi_2=\e^{-iy^2/2h}g_2(y)$, where now
%
%
\ba\label{eq_diff1}
\left (
\begin{array}{c}
g'_1 \\ g'_2
\end{array}
\right )
=\frac{\coupling}{h}
\left (
\begin{array}{cc}
0 & i\e^{-iy^2/h} \\
-i\e^{iy^2/h} & 0
\end{array}
\right )
\left (
\begin{array}{c}
g_1 \\ g_2
\end{array}
\right )=
\frac{\coupling}{h}
U(y)
\left (
\begin{array}{c}
g_1 \\ g_2
\end{array}
\right )
\ea
where $U$ is a unitary matrix. We can notice that the product $U(y_1)U(y_2)$ is
diagonal, which makes easier the computation of any multiple products of $U(y)$
\ba\label{eq_U12}
U(y_1)U(y_2)=
\left (
\begin{array}{cc}
\e^{-iy_1^2/h+iy_2^2/h} & 0 \\
0 & \e^{iy_1^2/h-iy_2^2/h}
\end{array}
\right )
\ea
The solution of~\eref{eq_diff1} is given by a series of matrix ordered products
and multiple integrals~\cite{Lam:1998}
\ba\label{eq_prod}
\left (
\begin{array}{c}
g_1(y) \\ g_2(y)
\end{array}
\right )
=
\left (
1+\frac{\coupling}{h}\int_{\yy}^y\d y_1 U(y_1)+
\frac{\coupling^2}{h^2}\int_{\yy}^y\d y_1\int_{\yy}^{y_1}\d y_2 U(y_1)U(y_2)+
\cdots \right )
\left (
\begin{array}{c}
g_1(\yy) \\ g_2(\yy)
\end{array}
\right )
\ea
Using the property~\eref{eq_U12} and setting $\omega(y)=y^2$ ($\omega$ can be
a more general function of $y$ as we shall see later), one can write a transfer
or S-matrix between two points $\yy$ and $y>0$ on the axis, away from the
tunneling region
\ba\label{eq_S}
\left (
\begin{array}{c}
g_1(y) \\ g_2(y)
\end{array}
\right )
=
\left (
\begin{array}{ll}
t & s \\
\bar s & \bar t
\end{array}
\right )
\left (
\begin{array}{c}
g_1(\yy) \\ g_2(\yy)
\end{array}
\right )
\ea
with $t\bar t-s\bar s=1$ by conservation of probabilities. The matrix elements
are infinite sums of ordered integrals given by
\bb\label{eq_elements}\fl
t=1+\frac{\coupling^2}{h^2}\int_{\yy}^y\d
y_1\int_{\yy}^{y_1}\d y_2\e^{-i\omega(y_1)/h+i\omega(y_2)/h} +
\cdots
\\ \nn\fl
s=\frac{\coupling}{h}\int_{\yy}^y\d y_1
\e^{-i\omega(y_1)/h}+
\frac{\coupling^3}{h^3}\int_{\yy}^y\d y_1\int_{\yy}^{y_1}\d y_2
\int_{\yy}^{y_2}\d y_3
\e^{-i\omega(y_1)/h+i\omega(y_2)/h-i\omega(y_3)/h}+\cdots
\ee
where the $y_i$ are dummy variables.
The characteristics of this matrix have been studied by many
authors~\cite{Rojo:2010,Kholodenko:2012} in the
case of the Zener effect~\cite{Rosen:1932}. In the
Gaussian case, when $\omega(y)$ is quadratic, it is convenient to use the theta
function representation in the complex
plane~\cite{Rojo:2010} when $y=\infty$.
Indeed the diagonal matrix element $t=\bar t$ can then be computed with the aid
of simple translation transformations.
For example, the double integral in the first line of~\eref{eq_elements}
can be simplified by introducing $\theta(x)=\oint \frac{\d
z}{2i\pi(z-i\epsilon)}\e^{izx}$, where the path in located on the
upper half complex plane, to satisfy the constraint $y_1>y_2$
\bb\nn\fl
\int_{-\infty}^{\infty}\d
y_1\int_{-\infty}^{y_1}\d y_2\e^{-i\omega(y_1)/h+i\omega(y_2)/h}
=
\int_{-\infty}^{\infty}\d
y_1\int_{-\infty}^{\infty}\d
y_2\oint\frac{\d z}{2i\pi}
\frac{\e^{-i\omega(y_1)/h+i\omega(y_2)/h+i(y_1-y_2)z}}{z-i\epsilon}
\\
=(\pi h)\oint\frac{\d z}{2i\pi}
\frac{\e^{ihz^2/4}}{z-i\epsilon}=\frac{\pi h}{2}
\ee
The last integral is obtained after translating $y_1\rightarrow y_1+hz/2$ and
$y_2\rightarrow y_2-hz/2$ respectively, to remove the couplings with $z$. Then
$t=1+\frac{\pi\coupling^2}{2h}+\cdots$. All the terms in
the series can be computed similarly, and the resummation leads to $t=
\e^{\pi\coupling^2/2h}$. We will introduce in the following the breakdown field
$h_0=\pi\coupling^2$ which is characteristic of the tunneling process. The same
techniques could be applied for elements $s$,
but one finds that the result is diverging in the large $y$ limit. The reason
is that the
phase of $s$ is diverging logarithmically~\cite{Torosov:2011}, as we  will see
below, although the modulus is finite. A correct asymptotic analysis for finite
$y$ and $\yy$ is therefore needed.
%
\subsection{Asymptotic analysis}
%
One can solve the equation for $g_1$ and $g_2$ using standard techniques.
Indeed, the differential equation satisfied by $g_1$ can be obtained,
separating $g_1$ from $g_2$ in~\eref{eq_diff1}
\bb\label{eq_diff_second}
g_1''+\frac{2iy}{h}g_1'=\frac{\coupling^2}{h^2}g_1,\;\;
g_2=\frac{h}{i\coupling}\e^{iy^2/h}g_1'
\ee
The two odd and even solutions for $g_1$ are a combination of two Kummer
fonctions $M$~\cite{Holmes} with an imaginary variable, and which can be chosen
such that
\bb\label{solg1}
g_1(y)=AM\left
(\frac{i\coupling^2}{4h},\frac{1}{2},-\frac{iy^2}{h}\right )+ByM\left (\frac{1}{
2}+\frac{i\coupling^2}{4h},\frac{3}{2},-\frac{iy^2}{h}\right )
\ee
where $A$ and $B$ are constant. Then $\varphi_1=\e^{iy^2/2h}g_1$ and
$\varphi_2=\e^{-iy^2/2h}g_2$. We notice that there are only two constants in the
problem, since from~\eref{eq_diff_second} $g_2$ is entirely determined by $g_1$.
The S-matrix~\eref{eq_S} between points $-y$ and $y>0$ can then be obtained by
eliminating the coefficients $A$ and $B$ in~\eref{solg1}. Setting
\bb\nn
g_1(\pm y)=Aa_1\pm Bb_1,\;g_2(\pm y)=\pm Aa_2+ Bb_2,
\ee
one can express the outgoing wavefunction $g_1(-y)$ and $g_2(y)$ as function of
an incoming wavefunction $g_1(y)$ and $g_2(-y)$ as represented
locally in~\efig{fig0}
\ba\label{eq_M}
\left (
\begin{array}{c}
g_1(-y) \\ g_2(y)
\end{array}
\right )
=
\left (
\begin{array}{ll}
1/t & -s/t \\
\bar s/t & 1/t
\end{array}
\right )
\left (
\begin{array}{c}
g_1(y) \\ g_2(\yy)
\end{array}
\right )
=
M
\left (
\begin{array}{c}
g_1(y) \\ g_2(\yy)
\end{array}
\right )
\ea
The functions $(a_1,a_2,b_1,b_2)$ depending on $y$ are given by
Kummer functions
\bb \nn \fl
a_1=M\left (\frac{i\coupling^2}{4h},\frac{1}{2},-\frac{iy^2}{h}\right ),\;
b_1=y M\left
(\frac{1}{2}+\frac{i\coupling^2}{4h},\frac{3}{2},-\frac{iy^2}{h}\right ),
\\ \nn\fl
a_2=-\frac{2y^2}{3\coupling}\e^{iy^2/h}\left (1+\frac{i\coupling^2}{2h}\right )
M\left (\frac{3}{2}+\frac{i\coupling^2}{4h},\frac{5}{2},-\frac{iy^2}{h}\right )
+\frac{h}{i\coupling}\e^{iy^2/h}M\left
(\frac{1}{2}+\frac{i\coupling^2}{4h},\frac{3}{2},-\frac{iy^2}{h}\right ),
\\ \fl
b_2=-y\frac{i\coupling}{h}\e^{iy^2/h}M\left (1+\frac{i\coupling^2}{4h},\frac{3
}{2},-\frac{iy^2}{h}\right ),
\ee
and the expression for the S-matrix elements is given by
\bb\nn
t=\bar t=\frac{a_1a_2+b_1b_2}{a_1a_2-b_1b_2},\;s=\frac{2a_1b_1}{a_1a_2-b_1b_2},
\;
\bar s=\frac{2a_2b_2}{a_1a_2-b_1b_2},
\;
t^2-s\bar s=1
\ee
Asymptotically, for $y$ large, one can use the expansion
$M(a,b,z)\simeq
\frac{\Gamma(b)}{\Gamma(b-a)}(-z)^{-a}+\frac{\Gamma(b)}{\Gamma(a) }\e^zz^{a-b}
$~\cite{Abramowitz} and keep the dominant terms
\bb\fl
g_1(\pm y)\simeq \sqrt{\pi}\left (\frac{iy^2}{h}\right )^{-i\coupling^2/4h}
\left (
\frac{A}{\Gamma(\ff-i\coupling^2/4h)}\pm
\frac{\sqrt{h}}{2\sqrt{i}} \frac{B}{\Gamma(1-i\coupling^2/4h)} \right )
\ee
and
\bb\fl
g_2(\pm y)\simeq \sqrt{\pi}
\left (\frac{-iy^2}{h}\right
)^{i\coupling^2/4h}
\left (
\pm
\frac{\coupling}{2\sqrt{ih}}
\frac{A}{\Gamma(1+i\coupling^2/4h)}-
\frac{ih}{\coupling}
\frac{B}{\Gamma(\ff+i\coupling^2/4h)}
\right )
\ee
Using the different duplication
formulas for gamma's functions: $\Gamma(\ff+ix)\Gamma(\ff-ix)=\pi/\cosh(\pi x)$,
$\Gamma(ix)\Gamma(1-ix)=\pi/i\sinh(\pi x)$, and
$\Gamma(\ff+ix)\Gamma(ix)=\sqrt{\pi}2^{1-2ix}\Gamma(2ix)$, one obtains
the probability of tunneling $p=1/t=\e^{-\pi\coupling^2/2h}=\e^{-h_0/2h}$, which
is the typical tunneling
amplitude already obtained in many previous
works~\cite{Slutskin:1967,Chambers:1968}. The breakdown field is
in this case equal to $h_0=\pi\coupling^2$ and corresponds exactly to the
semi-classical expression (see text further below). The remaining elements of
the tunneling matrix $M$ can be obtained after some algebra and one finds the
unitary matrix
\ba\label{eq_Mbis}
M=
\left (
\begin{array}{ll}
p & -iq\e^{-i\phi} \\
-iq\e^{i\phi} & p
\end{array}
\right )
\ea
where $q=\sqrt{1-p^2}$ and the phase $\phi$ depends on the coordinate $y$
\bb\label{phiy}
\phi(y)=-\frac{\pi}{4}+\frac{\coupling^2}{2h}\log\left
(\frac{2y^2}{h}\right )-\textrm{arg}\;
\Gamma(i\coupling^2/2h)
\ee
The phase diverges logarithmically with $y$.
Since the FS is not accounted for by~\efig{fig1}
for $|k_x|\gg 1$ where it should be more curved, we
assume that the phase is finite far from the tunneling region. Using a
Stirling
expansion of the gamma function in~\eref{phiy}, one finds that $\phi$ is finite
asymptotically only when
$y^2=h_0\e^{-1}/4\pi$. This corresponds approximately to
the coordinate where the tunneling region ends,  e.g. $y\simeq
\coupling$. In this case, instead of~\eref{phiy}, the phase is given by the
following regularization~\cite{Slutskin:1967,Kochkin:1968}
\bb\label{phi}
\phi=-\frac{\pi}{4}+u\log u-u-\textrm{arg}\Gamma(iu),\;\;u=\frac{h_0}{2\pi h}
\ee
The phase is zero in the low field limit ($u$ large) and equal to $\pi/4$ when
$h$ is large ($u$ small).
%
\section{Transmission through the small pocket}
%
\begin{figure*}
\centering
\includegraphics[width=0.8\textwidth, clip,angle=0]{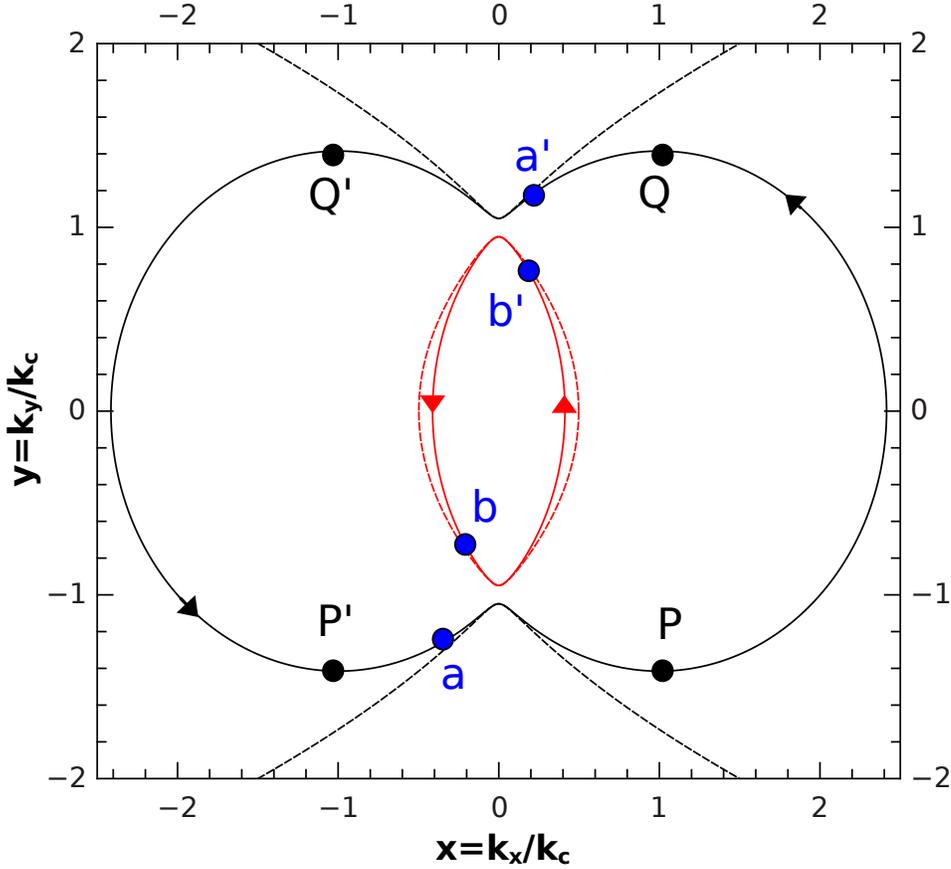}
\caption{
Effective two-band model. The dashed lines are the approximation
\eref{model_2bis} for small $x$. The parameters are $y_0=1$ and
$\coupling=0.05$. The shape of the small
lens, corresponding to the $\alpha$-orbit in magnetic field, is
slightly changed by the approximation when $y_0$ is small enough.}
\label{fig2}
\end{figure*}
%
A more general model is given by an hybridization of two parabolic bands,
whose Fermi surface is composed of two circular sheets, each of radius $k_0$
and centers $\pm k_c$, as displayed in \efig{fig2}, and for which the
Hamiltonian reads
\ba
\hat H
=
\left (
\begin{array}{cc}
\ff(k_x+k_c)^2+\ff(k_y^2-k_0^2) & \coupling \\
\coupling & \ff(k_x-k_c)^2+\ff(k_y^2-k_0^2)
\end{array}
\right )
\ea
Rescaling the variables with $k_c$ and setting
$x=k_x/k_c$, $y=k_y/k_c$, $\coupling/k_c^2\rightarrow \coupling$, and
$y_0^2=k_0^2/k_c^2-1>0$, one obtains
\ba\label{model_2}
\hat H
=
\left (
\begin{array}{cc}
\ff(x+1)^2+\ff(y^2-y_0^2-1) & \coupling \\
\coupling & \ff(x-1)^2+\ff(y^2-y_0^2-1)
\end{array}
\right )
\ea
For small $x$, one has the approximation near the tunneling points (points $a$,
$b$, $a'$, and $b'$ in~\efig{fig2})
\ba\label{model_2bis}
\hat H
\simeq
\left (
\begin{array}{cc}
x+\ff(y^2-y_0^2) & \coupling \\
\coupling & -x+\ff(y^2-y_0^2)
\end{array}
\right )
\ea
This Hamiltonian gives a first order differential matrix equation, similar
to~\eref{eq_diff1}, after setting $\varphi_1(y)=\e^{i\omega(y)/2h}g_1(y)$ and
$\varphi_2(y)=\e^{-i\omega(y)/2h}g_2(y)$
\ba\label{eq_diff2bis}
\left (
\begin{array}{c}
g'_1 \\ g'_2
\end{array}
\right )
=\frac{\coupling}{h}
\left (
\begin{array}{cc}
0 & i\e^{-i\omega(y)/h} \\
-i\e^{i\omega(y)/h} & 0
\end{array}
\right )
\left (
\begin{array}{c}
g_1 \\ g_2
\end{array}
\right )
\ea
with $\omega(y)=(y^3/3-y_0^2y)$ instead of $\omega(y)=y^2$. The first
double integral in~\eref{eq_elements} contributing to $t$ in the large field
limit and far from the scattering region can be written as
\bb\fl
\int_{-\infty}^{\infty}\d y_1\int_{-\infty}^{y_1}\d
y_2\e^{-i\omega(y_1)/h+i\omega(y_2)/h}
=\int_{-\infty}^{\infty}\d
y_1\int_{-\infty}^{\infty}\d y_2
\oint \frac{\d
z\e^{-i\omega(y_1)/h+i\omega(y_2)/h+iz(y_1-y_2)}}{2i\pi(z-i\epsilon)}
\ee
We can define each integral over $y_1$ and $y_2$ as a function of $z$
\bb
h(z)=\int_{-\infty}^{\infty}\d y\e^{-i\omega(y)/h+izy}=2\pi h^{1/3}\Ai\left [
-h^{1/3}\left ( \frac{y_0^2}{h}+z \right )\right ]
\ee
Then using $(z-i\epsilon)^{-1}=P(1/z)+i\pi\delta(z)$, one obtains
\bb\label{t_2bis}
t\simeq 1+\frac{\coupling^2}{h^2}\left [
2\pi^2 h^{2/3}\Ai^2\left (
-h^{1/3}\frac{y_0^2}{h}\right )+\frac{1}{2i\pi}\int_0^{\infty}\frac{\d
z}{z}[h^2(z)-h^2(-z)] \right ]
\ee
This expression is valid at large fields. It contains an imaginary part which is
due to the presence of the small $\alpha$-orbit between points $b$ and
$b'$, with area $S_{\alpha}$, in red in~\efig{fig2}. Indeed, after tunneling
through $a$, the particle can be scattered multiple times around the $\alpha$ orbit, and therefore acquires a phase proportional to $S_{\alpha}$, before
exiting trough $a'$.
%
\begin{figure*}
\centering
\includegraphics[width=0.8\textwidth, clip,angle=0]{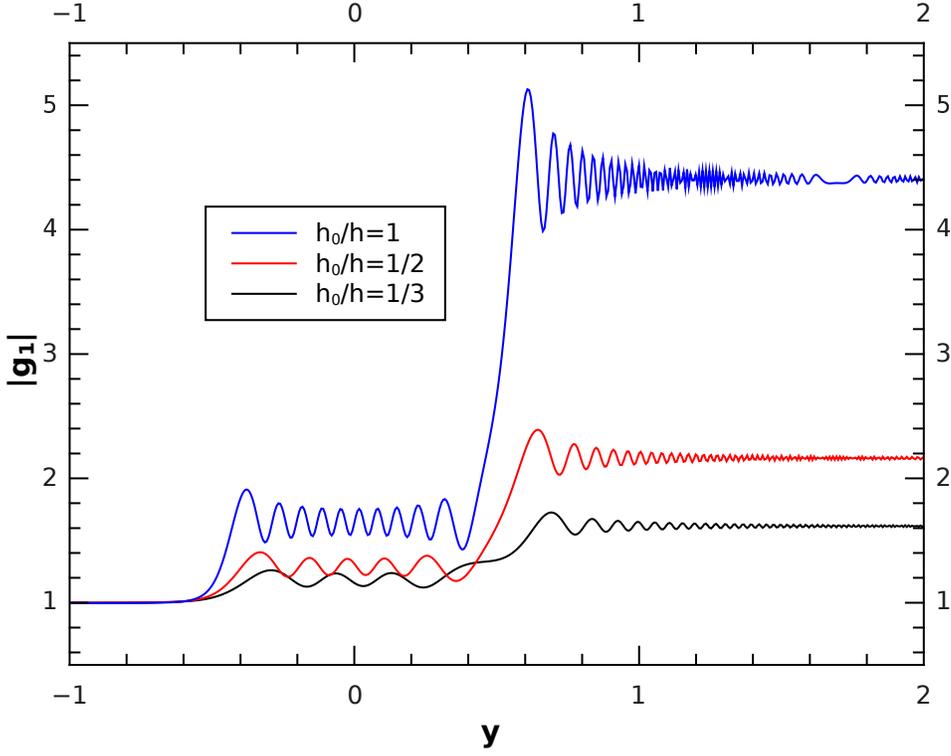}
\caption{Wave profile of $g_1$ as function of $y$ for three different values of
the inverse field ratio $h_0/h$. Parameters are $y_0=0.5$ and $\coupling=0.02$.
From the initial condition $g_1(y\ll -1)=1$, we have
integrated~\eref{eqdiffH}. The ratio between the two amplitudes $g_1(y\gg
1)/g_1(y\ll -1)$ is proportional to the inverse of
tunneling probability $\e^{h_0/h}=1/p^2$, up to some oscillation
factor which corresponds to
interferences in the $\alpha$-pocket (see text). Indeed the electron has to
cross two breakdown regions, therefore a factor $p^2$ is involved.}
\label{fig4}
\end{figure*}
In the following we compare the
transmission coefficient $T=1/|t|^2$ through the small $\alpha$-orbit to
the expression given by the semi-classical relation and numerical results.
%
\subsection{Semiclassical approximation}
%
The Hamiltonian~\eref{model_2bis} leads to the set of differential equations
for $g_1$ and $g_2$
\bb\label{eqdiffH}
h^2g_1''+ih\omega'(y)g_1'-\coupling^2g_1=0,\;
h^2g_2''-ih\omega'(y)g_2'-\coupling^2g_2=0
\ee
with $\omega'(y)=y^2-y_0^2$~\footnote{The solutions of~\eref{eqdiffH} are
actually given by triconfluent Heun functions~\cite{Hortacsu:2012}}.
In~\efig{fig4}, we have represented the
numerical solution of~\eref{eq_diff2bis} and~\eref{eqdiffH}, in particular the
modulus of $|g_1|$ for different values of fields. At large values of $y$, we
can
approximate~\eref{eqdiffH} by the equations $ihy^2g_1'-\coupling^2g_1\simeq 0$
and $ihy^2g_2'+\coupling^2g_2\simeq 0$, which leads to $g_1\simeq
\e^{i\coupling^2/(hy)}\sim$ constant, and $g_2\simeq
\e^{-i\coupling^2/(hy)}\sim $ constant. We have chosen $g_1(y\ll -1)=1$ and
integrated numerically the first differential equation. On the far right, $y\gg
1$, the constant value is proportional to $\e^{h_0/h}=1/p^2$. Therefore, by
computing $t$, we can access to the breakdown field $h_0$.
The semi-classical approximation
$g_1(y)=\exp(iS(y)/h)$, where $S$ corresponds physically to an area enclosed
by the trajectory, consists in expanding $S(y)$ as a series in $h\ll 1$. In
particular, at the leading order in $h$ for small field values, one can
write $S=S_0+hS_1+\cdots$ with
\bb\nn
S_0'^2+\omega'(y)S_0'+\coupling^2=0,
\\
S_0'=\frac{1}{2}\left (
-\omega'(y)\pm \sqrt{\omega'(y)^2-4\coupling^2} \right )
\ee
When $\omega(y)=y^2$, as for the model~\eref{model1} (linear sheets of
\efig{fig1}), $S_0(y)=-y^2/2\pm
\ff y^2\sqrt{y^2-\coupling^2}\mp \ff\coupling^2\log(y+\sqrt{y^2-\coupling^2})$.
The breakdown field $h_0$ is then given by the tunneling amplitude
$p=\exp(-h_0/2h)$ through the forbidden region, or
$h_0=2\int_{-\coupling}^{\coupling}\sqrt{\coupling^2-y^2}=\coupling^2\pi$,
which corresponds
to the exact result in this particular case. For the second
model,~\eref{model_2bis}  (parabolic sheets of \efig{fig2}), the breakdown
field through one of the two tunneling
regions, is instead given by
\bb
h_0=\int_{\sqrt{y_0^2-2\coupling}}^{\sqrt{y_0^2+2\coupling}}
\sqrt{4\coupling^2-(y^2-y_0^2)^2}\d y\simeq \frac{\pi\coupling^2}{y_0}
\ee
The phase variation of $S_0$ around the small pocket corresponds to the
area $S_{\alpha}$ of the pocket
\bb\nn
S_{\alpha}=2\int_0^{\sqrt{y_0^2-2\coupling}}\sqrt{(y^2-y_0^2)^2-4\coupling^2}\d
y
\\
=\frac{4}{3}\sqrt{y_0^2+2\coupling}\left [
y_0^2E\left (\sqrt{\frac{y_0^2-2\coupling}{y_0^2+2\coupling}}\right )
-2\coupling K\left (\sqrt{\frac{y_0^2-2\coupling}{y_0^2+2\coupling}}\right )
\right ]\simeq \frac{4}{3}y_0^3
\ee
where $E$ and $K$ are complete elliptic functions of the second and first kind
respectively, and the approximation is taken when $\coupling$ is small.
For a unit cell parameter $a=10$\AA, or, equivalently, a unit
cell area of 100\AA$^2$, which holds for the organic metals
$\theta$-(ET)$_4$CoBr$_4$(C$_6$H$_4$Cl$_2$) and $\kappa$-(ET)$_2$Cu(SCN)$_2$,
the frequency $F_{\alpha}$ and magnetic breakdown field $B_0$, expressed in
Tesla are given by
\bb
F_{\alpha}=\frac{2\pi \hbar S_{\alpha}}{a^2e}=4136\,S_{\alpha}[\textrm{T}],
\; B_0=\frac{(2\pi)^2 \hbar}{a^2e}h_0=25\,988\,h_0[\textrm{T}]
\ee
As examples, the frequency $F_{\alpha}$ of the two above salts
is 944 T and 600 T, respectively, yielding $y_0$ = 0.55 and 0.48. The MB field
$B_0$ is 35 T and 16 T, yielding $\coupling$ = 0.015 and 0.01, respectively.
%
\subsection{Transmission coefficient}
%
We consider the probability of tunneling between points $P$ and $Q$
in~\efig{fig2}, using the model~\eref{model_2bis}, which is defined
by the modulus $T=|\varphi_1(Q)/\varphi_1(P)|^2=1/|t|^2$.
Given the approximate value of $t$ in~\eref{t_2bis}, we can estimate $T$ in
the large field limit by exponentiating~\eref{t_2bis}

\begin{figure*}
\centering
\includegraphics[width=0.8\textwidth, clip,angle=0]{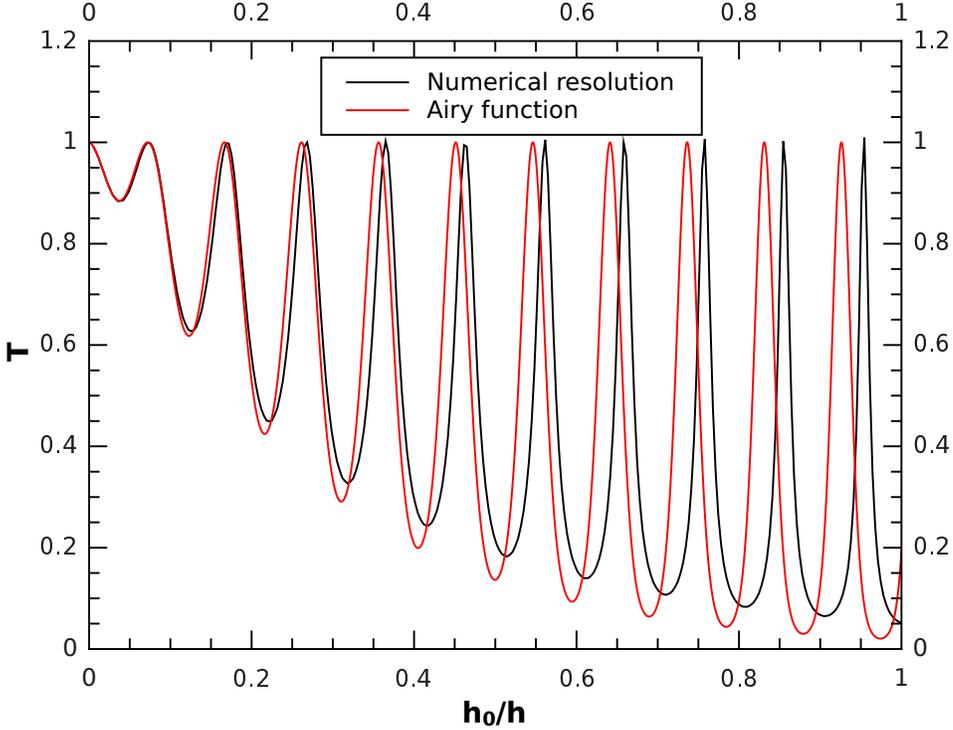}
\caption{Transmission coefficient as a function of the inverse field $h_0/h$
for $y_0=0.5$ and a hybridization coupling $\coupling=0.02$ ($h_0=0.002513$
and $S_{\alpha}=0.159598$). The black line are computed by
solving the
differential equations~\eref{eq_diff2bis} and the red line is the large field
approximation~\eref{TappAiry} obtained by computing approximately $t$ in the
S-matrix~\eref{t_2bis}.}
\label{fig3Airy}
\end{figure*}

\bb\label{TappAiry}
T\simeq \exp\left [-\frac{4\pi^2\coupling^2}{h^{4/3}}\Ai^2\left
(-h^{1/3}\frac{y_0^2}{h}\right )\right ]
\ee
$T$ reaches its maximum, or resonance value $T=1$, whenever the Airy function
vanishes. This happens when $h=y_0^3(-a_n)^{-3/2}$, where $a_n<0$ are the zeroes
of the Airy functions. For example $a_1=-2.33811$, $a_2=-5.08795$. A
comparison with the numerical resolution of the differential
equations~\eref{eq_diff2bis} is shown in \efig{fig3Airy}. The approximation
presents a phase shift more pronounced as the field decreases.
%
%
Semi-classically, we can compute $T$ using the tunneling
matrix~\eref{eq_Mbis} between the two points $P$ and $Q$ in~\efig{fig2}.
It is the contribution of all possible trajectories between the two points,
including the multiple reflections inside the $\alpha$-orbit
\bb\fl\nn
\varphi_1(Q)=\varphi_1(P)(pi\e^{iS_{\alpha}/2h}p)
+\varphi_1(P)[pi\e^{iS_{\alpha}/2h}(-q\e^{-i\phi})i\e^{iS_{\alpha}/2h}
(-q\e^{-i\phi})i\e^{iS_{\alpha}/2h}p]+\cdots
\\
=\varphi_1(P)\frac{ip^2\e^{iS_{\alpha}/2h}}{1+q^2\e^{iS_{\alpha}/h-2i\phi}}
\ee
The factor $i$ corresponds to passing each of the two singular (or
turning) points on the surface $\alpha$~\efig{fig2} where the slopes are
infinite. The phase $\phi$ is taken from~\eref{phi}. Therefore one obtains
(see~\cite{Kaganov:1983})
\bb\label{Tsem}
T=\frac{p^4}{1+q^4+2q^2\cos(S_{\alpha}/h-2\phi)}
\ee
$T$ is maximum when the field satisfies
$\cos(S_{\alpha}/h-2\phi)=-1$, e.g. $T=1$, and the quantized values are given by
\bb
h=\frac{S_{\alpha}}{2\pi n+\pi+2\phi(h)}
\ee
If $\phi\simeq \pi/4$, then $h_0/h=3\pi h_0/2S_{\alpha},7\pi
h_0/2S_{\alpha},\cdots$.
%
\begin{figure*}
\centering
(a)
\includegraphics[width=0.8\textwidth, clip,angle=0]{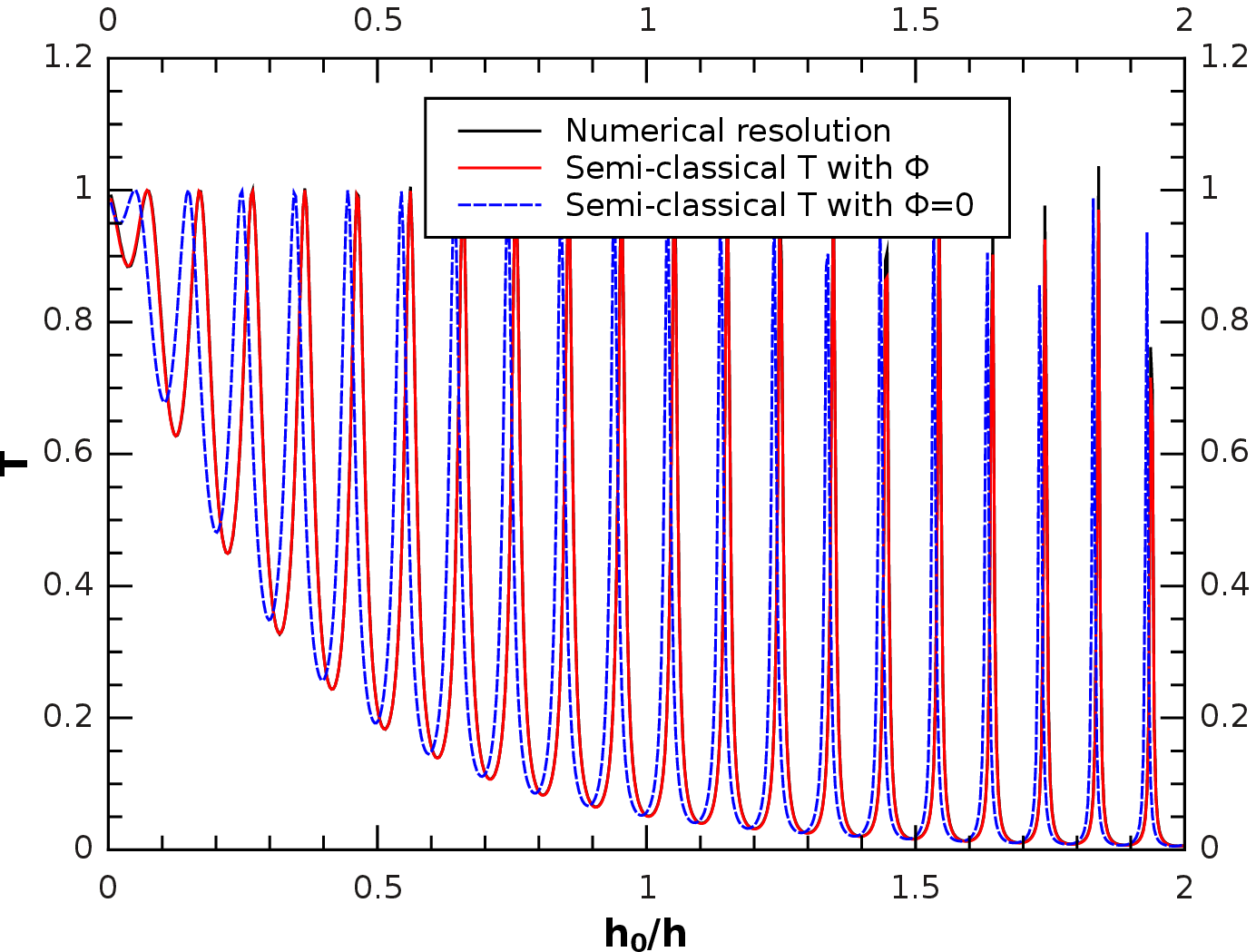}
(b)
\includegraphics[width=0.8\textwidth, clip,angle=0]{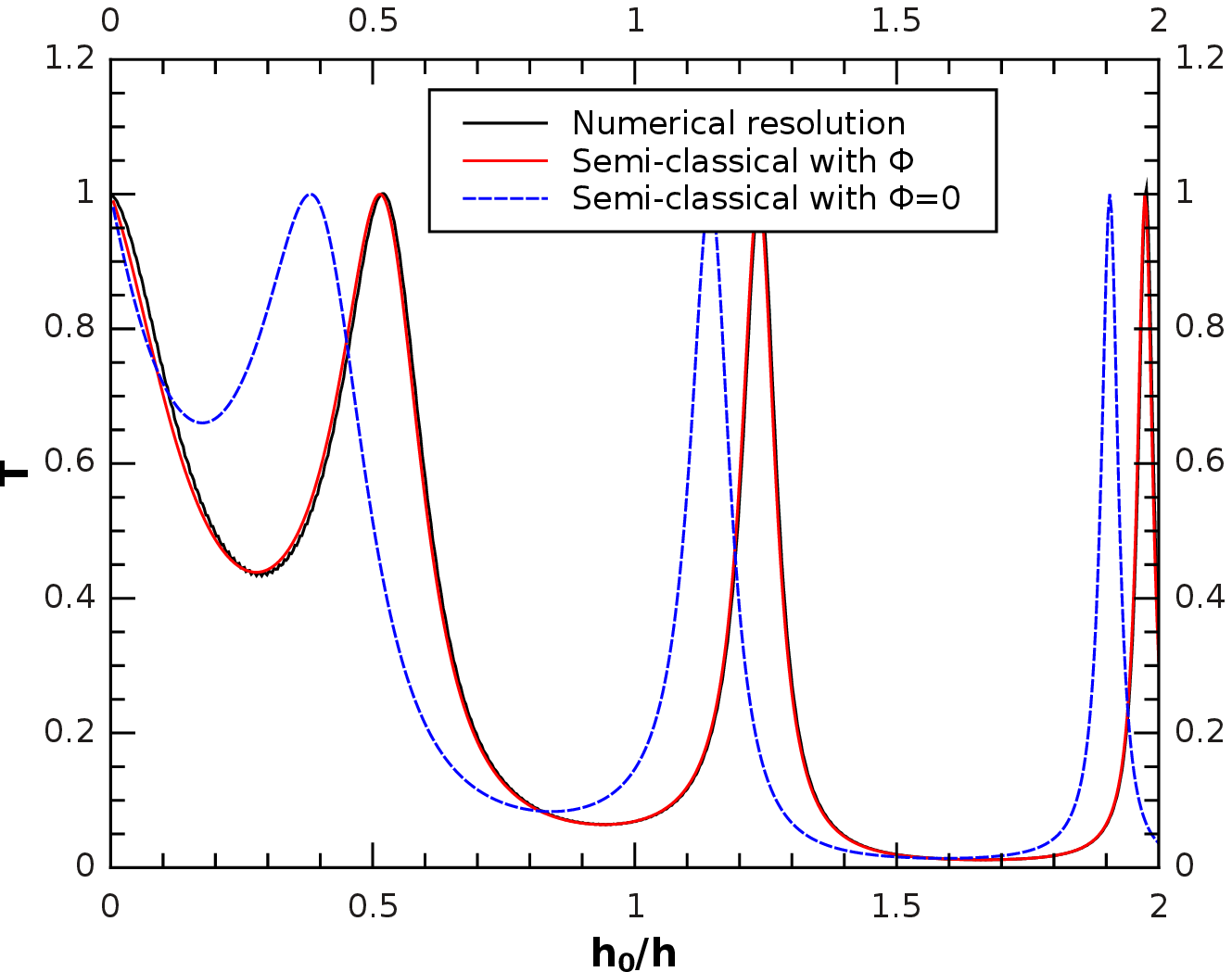}
\caption{
Transmission coefficient as a function of the inverse field $h_0/h$ for $y_0=0.5$ and for two
values of hybridization coupling: (a)  $\coupling=0.02$, $h_0=0.002513$,
$S_{\alpha}=0.159598$ and (b) $\coupling=0.05$, $h_0=0.015959$ and
$S_{\alpha}=0.131460$). Black lines are computed by solving
the differential equations~\eref{eq_diff2bis}. Red lines, which are indiscernible from
the black lines, are the
result of~\eref{Tsem} where the phase $\phi$ is given by~\eref{phi}. The dotted lines are obtained without reflection phase ($\phi=0$). $\phi=0$ only holds in the limit of small fields ($h_0/h \gg 1$).}
\label{fig3}
\end{figure*}
%
In~\efig{fig3} is plotted the transmission coefficient as function of the
inverse field $h_0/h$. The black continuous lines are obtained by
solving the system of differential equations~\eref{eq_diff2bis}, with the
condition $g_1(-y_c)=1$, $g_2(-y_c)=0$, $y_c=5$, then by computing
the ratio $T=1/|t|^2=|g_1(-y_c)/g_1(y_c)|^2$. Without the phase $\phi$ from the
reflection coefficient~\eref{phi}, the values differ increasingly as the field
is
increased (dotted blue lines). Oppositely, the phase does not contribute to
the oscillations when the field becomes small.
%
\section{Amplitude ratios between two-interacting orbits}
%
In this section, we consider the model~\eref{model_2}, which represents the
hybridization of the two giant orbits corresponding to the $\beta$-orbit of the organic metals considered in the last section (see \efig{fig0}). Using the field quantization, one obtains
the set of differential equations
\bb\nn
-h^2\partial^2_y\varphi_1+2ih\partial_y\varphi_1+(y^2-y_0^2)\varphi_1
+2\coupling\varphi_2=0,
\\ \label{eq_diff2}
2\coupling\varphi_1-h^2\partial^2_y\varphi_2-2ih\partial_y\varphi_2+(y^2-y_0^2)
\varphi_2=0
\ee
As in preceding sections, we introduce two functions $g_1$ and $g_2$ such
that $\varphi_i(y)=g_i(y)\exp(i\omega_i(y)/h)$. $\omega_i$ are two phase
functions
that are chosen such that the coefficient of $g_i$ vanishes in~\eref{eq_diff2}
after replacement. One obtains
\bb\nn
-h^2g_1''-2ih(\omega_1'-1)g_1'+2\coupling g_2\exp[i(\omega_2-\omega_1)/h]=0,
\\ \label{eq_diff2_g}
-h^2g_2''-2ih(\omega_2'+1)g_2'+2\coupling g_1\exp[i(\omega_1-\omega_2)/h]=0
\ee
The phase functions satisfy the differential equations
\bb \label{eq_diff_omega}
\fl
-ih\omega_1''+\omega_1'^2-2\omega_1'+y^2-y_0^2=0,
\;\textrm{and}
-ih\omega_2''+\omega_2'^2+2\omega_2'+y^2-y_0^2=0
\ee
We can chose in particular $\omega_1=\omega$ and $\omega_2=-\bar\omega$.
The solutions of the Ricatti equations with respect to
$\omega'$ defined by~\eref{eq_diff_omega} can
be found in principle using hypergeometric functions.
The coefficients $\omega_1'-1$ and $\omega_2'+1$ in front of the $g_i'$s
in~\eref{eq_diff2_g} can be removed using an additional transformation
$g_i'(y)=h_i(y)\exp(2i\theta_i(y)/h)$, such that
\bb
\theta_1=y-\omega_1,\;\theta_2=-y-\omega_2
\ee
Then finally
\bb\nn
h_1'=\frac{2\coupling}{h^2}g_2\e^{-2iy/h+i(\omega_1+\omega_2)/h},
\;
h_2'=\frac{2\coupling}{h^2}g_1\e^{2iy/h+i(\omega_1+\omega_2)/h},
\ee
The whole system can be cast into a system of first-order differential equations
\ba\label{eq_diff2_sys}
\left (
\begin{array}{c}
g'_1 \\ g'_2 \\ h_1' \\ h_2'
\end{array}
\right )
=
\left (
\begin{array}{cc}
0 & V \\
U & 0
\end{array}
\right )
\left (
\begin{array}{c}
g_1 \\ g_2 \\ h_1 \\ h_2
\end{array}
\right )
\ea
with $U$ and $V$ defined by
\ba
U=\frac{2\coupling\e^{i(\omega_1+\omega_2)/h}}{h^2}
\left (
\begin{array}{cc}
0 & \e^{-2iy/h}
\\ \e^{2iy/h} & 0
\end{array}
\right ),\;
V=\left (
\begin{array}{cc}
\e^{2iy/h-2i\omega_1/h} & 0
\\
0 & \e^{-2iy/h-2i\omega_2/h}
\end{array}
\right )
\ea
The S-matrix can then be formally defined by ordered-integral iterations of the
matrix functions $U$ and $V$, similarly as~\eref{eq_U12}. If we introduce
$u(y)=\exp(2iy/h-2i\Im(\omega)/h)$ and
$v(y)=\exp(2iy/h-2i\omega/h)$, one finds that the $t$ matrix element can be
expanded as
\bb\nn
t=1+\frac{4\coupling^2}{h^4}\int_{y\ge y_1\ge y_2 \ge y_3 \ge y_4 \ge -y}
v(y_1)\bar u(y_2)\bar v(y_3) u(y_4)
\\
+\frac{16\coupling^4}{h^8}\int_{y\ge y_1\ge \cdots \ge y_8 \ge -y}v(y_1)\bar
u(y_2)\bar v(y_3) u(y_4)
v(y_5)\bar u(y_6)\bar v(y_7) u(y_8)+\cdots
\ee
which is equivalent to~\eref{eq_elements} found for one tunneling junction.
%
\subsection{Case with no hybridization ($\coupling=0$)}
\begin{figure*}
\centering
\includegraphics[width=0.5\textwidth, clip,angle=0]{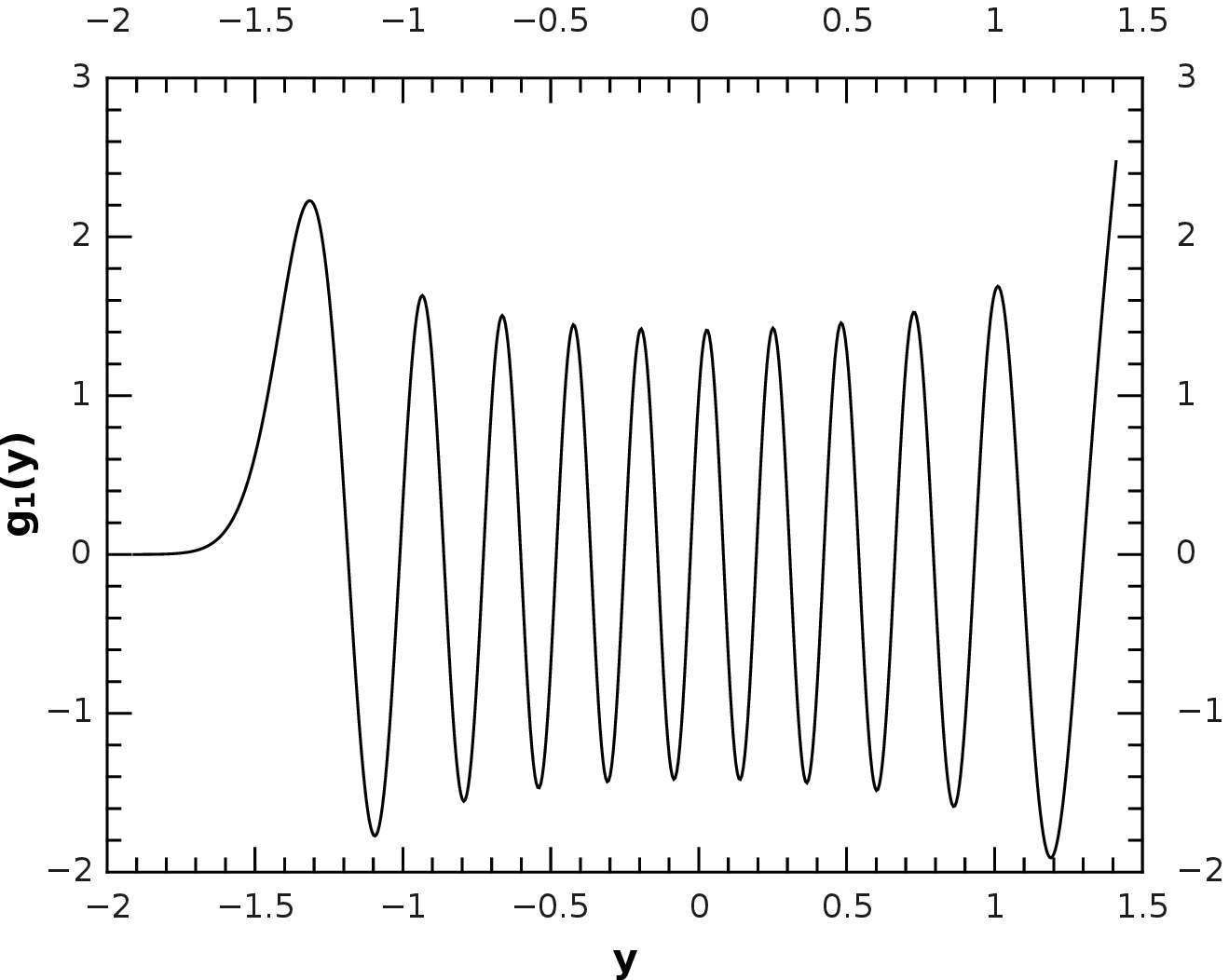}(a)
\includegraphics[width=0.5\textwidth, clip,angle=0]{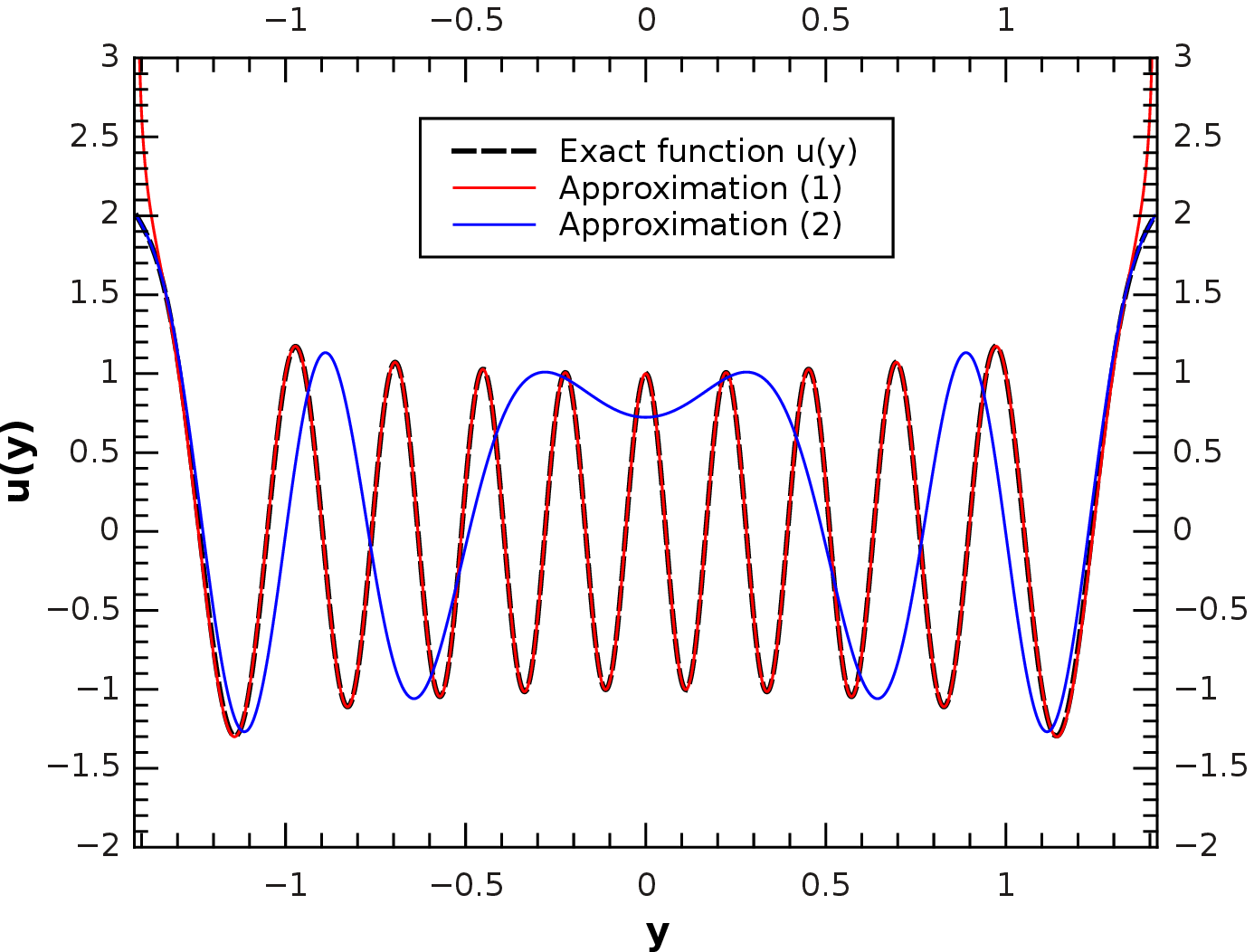}(b)
\includegraphics[width=0.5\textwidth, clip,angle=0]{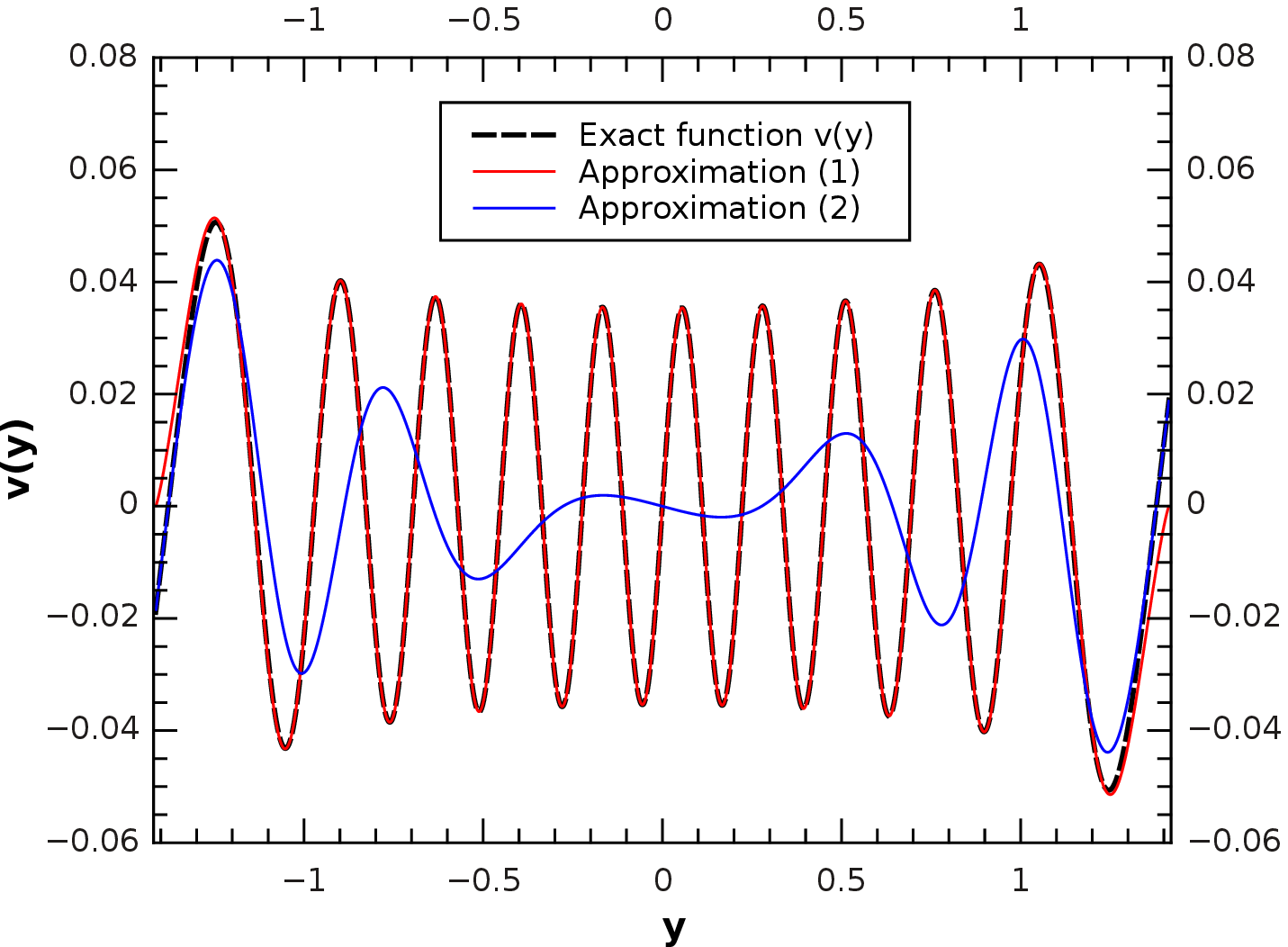}(c)
\caption{Wave profile of functions $g_1$ (a), $u$ (b), and $v$ (c) as a
function of $y$ for field value $h=0.05$ and parameters
$y_0=1$, $\coupling=0$ ($r^2=2$). Approximation (1) is given
by Eqs.~\ref{wavefunctions_appr_y_u} and~\ref{wavefunctions_appr_y_v}, which are
accurate in the bulk $-r<y<r$, and approximation (2) by
Eqs.~\ref{wavefunctions_appr_r_u} and~\ref{wavefunctions_appr_r_v}, which are
correct only near the borders of the
turning points $y=\pm r=\pm\sqrt{2}$. Function $g_1$ vanishes as $y\rightarrow
-\infty$ but is unbounded when $y\rightarrow\infty$.}
\label{fig5}
\end{figure*}

%
In absence of hybridization, it is interesting to study the phase for an
unbounded state (a state where one of the boundary condition for the
wavefunction does not vanish at infinity).
The two sheets decouple in this case, and
one has only two independent linear second-order differential equations for
$g_1$ and $g_2$.
Setting $\varphi_1=g_1(y)\e^{iy/h}$ and $\varphi_2=g_2(y)\e^{-iy/h}$,
~\eref{eq_diff2} becomes
\bb
h^2g_1''(y)=(y^2-r^2)g_1(y),\;\;h^2g_2''(y)=(y^2-r^2)g_2(y)
\ee
where $r^2=1+y_0^2$ is the radius of the $\beta$ orbit. It is well-known
that the even and
odd solutions are expressed using two Kummer functions $M$ with $y^2/h$
as main argument~\cite{Holmes}
\bb\nn
g_1(y)=A\e^{-y^2/2h}M\left
(\frac{1}{4}-\frac{r^2}{4h},\frac{1}{2},\frac{y^2}{h}\right)
+ByM\left
(\frac{3}{4}-\frac{r^2}{4h},\frac{3}{2},\frac{y^2}{h}\right)
\\
=A u(y)+B v(y)
\ee
Solution for the other function $g_2$ is similar with independent constants. We
impose the constraint that, for $y$ large and negative, $g_1$ vanishes.
This leads to the relation
\bb\label{constantsAB}
\frac{B}{2\Gamma\left (\frac{3}{4}-\frac{r^2}{4h}\right )}
-\frac{A}{\sqrt{h}\Gamma\left
(\frac{1}{4}-\frac{r^2}{4h}\right )}=0
\ee
In~\efig{fig5}(a) is represented $g_1$, with a vanishing boundary condition on
the left. Only one constant remains, which is not relevant when we consider the
ratio
of the wave function between $P$ and $Q$ in~\efig{fig2}. Indeed the transmission
factor defined here by $T=|g_1(-r)/g_1(r)|^2$ is exactly equal to
\bb\fl
T=\left |
\frac{
\Gamma\left (\frac{1}{4}-\frac{r^2}{4h}\right )
M\left (\frac{1}{4}-\frac{r^2}{4h},\frac{1}{2},\frac{r^2}{h} \right )
+\frac{2r}{\sqrt{h}}\Gamma\left (\frac{3}{4}-\frac{r^2}{4h}\right )
M\left(\frac{3}{4}-\frac{r^2}{4h},\frac{3}{2},\frac{r^2}{h}
\right )}
{\Gamma\left (\frac{1}{4}-\frac{r^2}{4h}\right )
M\left (\frac{1}{4}-\frac{r^2}{4h},\frac{1}{2},\frac{r^2}{h} \right )
-\frac{2r}{\sqrt{h}}\Gamma\left (\frac{3}{4}-\frac{r^2}{4h}\right )
M\left(\frac{3}{4}-\frac{r^2}{4h},\frac{3}{2},\frac{r^2}{h}
\right )}
\right |^2
\ee
and is a function of $r^2/h$. In physical units, the ratio $r^2/2h$ is equal to
the $\beta$-orbit frequency
(in Tesla) divided by the magnetic field $B$
\bb
\frac{r^2}{2h}=\frac{F_{\beta}}{B}
\ee
which is usually a large number ($F_{\beta}$ is few thousands of Tesla for organic conductors). It
has to be noticed that imposing a
vanishing wavefunction at both negative and positive large
values of $y$ (bound state) leads to two conditions
\bb
\frac{B}{2\Gamma\left (\frac{3}{4}-\frac{r^2}{4h}\right )}
\pm \frac{A}{\sqrt{h}\Gamma\left
(\frac{1}{4}-\frac{r^2}{4h}\right )}=0
\ee
which can only be satisfied when the gamma functions are infinite. This
happens when both arguments of the gamma functions are negative integers, and
one obtains the usual
quantification relation or Landau levels $r^2=(2n+1)h$ with $n$ positive
integer. Using the different asymptotic expansions
for the Kummer function~\cite{Abramowitz}, one obtains for each wave function
$u$ and $v$ a good approximation near the turning points $y\simeq \pm r$  (see
\efig{fig5}(b), and (c),
approximation (2))
\bb\nn
u(y)\simeq\sqrt{\pi}\left (\frac{r^2}{2h}\right )^{1/6}
\left \{
\Ai\left [ \left(\frac{r^2}{2h}\right )^{2/3}\left ( \frac{y^2}{r^2}-1\right
)\right ]
\cos \left ( \frac{\pi}{4}-\frac{\pi r^2}{4h} \right )
\right .
\\ \label{wavefunctions_appr_r_u}
+
\left .
\Bi\left [ \left(\frac{r^2}{2h}\right )^{2/3}\left ( \frac{y^2}{r^2}-1\right
)\right ]
\sin \left ( \frac{\pi}{4}-\frac{\pi r^2}{4h} \right )
\right \},
\\ \nn
v(y)\simeq \frac{\sqrt{\pi}}{2}\left (\frac{r^2}{2h}\right )^{-5/6}
y\left \{
\Ai\left [ \left(\frac{r^2}{2h}\right )^{2/3}\left ( \frac{y^2}{r^2}-1\right
)\right ]
\cos \left ( \frac{3\pi}{4}-\frac{\pi r^2}{4h} \right )
\right .
\\ \label{wavefunctions_appr_r_v}
+
\left .
\Bi\left [ \left(\frac{r^2}{2h}\right )^{2/3}\left ( \frac{y^2}{r^2}-1\right
)\right ]
\sin \left ( \frac{3\pi}{4}-\frac{\pi r^2}{4h} \right )
\right \}
\ee
In the region $-r<y<r$, not too close to the turning points, the solutions are
instead adequately approximated by (see \efig{fig5}(b) and (c), approximation
(1))
\bb
\label{wavefunctions_appr_y_u}
u(y)\simeq \frac{1}{\sqrt{\sin\theta}}\left [
\cos\left ( \frac{r^2}{2h}\left (\theta-\ff\sin 2\theta \right ) \right )
-\sin\left (\frac{\pi r^2}{4h} \right ) \right ],
\\ \label{wavefunctions_appr_y_v}
v(y)\simeq -\frac{h}{r\sqrt{\sin\theta}}\left [
\sin\left ( \frac{r^2}{2h}\left (\theta-\ff\sin 2\theta \right ) \right )
-\sin\left (\frac{\pi r^2}{4h} \right ) \right ]
\ee
%
%
Moreover, the ratio between the two constants $B$ and $A$ in~\eref{constantsAB}
is approximated by
\bb\label{constantsAB_appr}
\frac{B}{A}=\frac{2\Gamma\left (\frac{3}{4}-\frac{r^2}{4h}\right
)}{\sqrt{h}\Gamma\left
(\frac{1}{4}-\frac{r^2}{4h}\right )}
\simeq
\frac{r}{h}\cot\left (\frac{\pi r^2}{4h}+\frac{\pi}{4}\right )
\ee
Using \eref{wavefunctions_appr_r_u} and~\eref{wavefunctions_appr_r_v} for
$y=\pm r$, and $\Bi(0)/\Ai(0)=\sqrt{3}$,
one obtains the semi-classical
limit of the inverse transmission factor, and after some algebra and
simplifications one obtains the simple result
\bb\label{Tbeta}
T\simeq 4\sin^2\left (\frac{\pi r^2}{2h}+\frac{\pi}{3} \right )
=4\sin^2\left (\pi\frac{F_{\beta}}{B}+\frac{\pi}{3} \right )
\ee
The frequency of the oscillations is $F_{\beta}/2$ as expected, but there
is a shift equal to $\delta=\pi/3$ as opposed to the semi-classical limit,
which is equal to $\delta=\pi/2$ for a bound state or localized
wavefunction, where at each turning point a Maslov factor equal to
$\pi/2$ is involved after total reflection of the wave function.
%
\subsection{Semi-classical analysis for interacting orbits}
%
In this section, one computes semi-classically for a bound state the amplitude
ratio between points $P$ and $Q$ in~\efig{fig2}, using a transfer matrix method
to obtain all the contributions from the different electronic paths. One
has indeed to
evaluate the sum of all the amplitudes corresponding
to multiple orbits connecting the two points $P$ and $Q$, with their harmonics,
and using the connection formula~\eref{eq_Mbis} for the tunneling regions.
In~\efig{fig2}, we have represented 4 different points
(amplitudes) $(a,a',b,b')$. $a$ and
$a'$ belong to orbits $\beta$ or $2\beta-\alpha$, and $b$ and $b'$ belong
to orbits $\alpha$ or $\beta$. These points are located just before the
tunneling event, such that there is a possibility to be transmitted or
reflected, just
after passing trough the breakdown points. A trajectory is an ensemble of steps
on the
surface, which
connect $P$ to $Q$. At time $n=0$ we start from $P$. At later time $n+1$, we
can write the amplitudes as function of the amplitudes at time $n$. For example
amplitude $b$ at time $n+1$ is the sum of $b'$ after reflection and $a'$ after
tunneling at time $n$, and can be written as $b(n+1)=p
\e^{iS_{\alpha}/2h}a'(n)-q\e^{iS_{\alpha}/2h-i\phi}b'(n)$. There are 3 other
equations connecting the different points at each step on a trajectory. At $P$,
$Q$, $P'$ and $Q'$ we introduce a phase shift $\delta=\pi/2$. One can write
therefore the system
\bb\nn
a(n+1)=q\e^{i(S_{\beta}-S_{\alpha}/2)/h+i\phi+2i\delta}a'(n)+p\e^{i(S_{\beta}-S_
{ \alpha } /2)/h+2i\delta} b'(n)
\\ \nn
a'(n+1)=q\e^{i(S_{\beta}-S_{\alpha}/2)/h+i\phi+2i\delta}a(n)+p\e^{i(S_{\beta}-S_
{ \alpha }/2)/h+2i\delta}b(n)
\\ \nn
b(n+1)=-q\e^{iS_{\alpha}/2h-i\phi}b'(n)+p\e^{iS_{\alpha }/2h}a'(n)
\\
b'(n+1)=-q\e^{iS_{\alpha}/2h-i\phi}b(n)+p\e^{iS_{\alpha }/2h}a(n)
\ee
From these relations, we can define a step matrix $R$, acting on vector
$\bv(n)^\mathsf{T}=(a(n),b(n),a'(n),b'(n))$, with initial condition
$\bv(0)^\mathsf{T}=(0,0,\e^{-i(S_{\beta}-S_{\alpha}/2)/h-i\delta},0)$. Then
$\bv(n+1)=R\bv(n)$, with
\ba\label{step_matrix}
R=
\left (
\begin{array}{cc}
0 & A \\ A & 0
\end{array}
\right )
,\;
A=
\left (
\begin{array}{cc}
qx_{2\beta-\alpha}\e^{i\phi} & px_{2\beta-\alpha} \\
px_{\alpha} & -qx_{\alpha}\e^{-i\phi}
\end{array}
\right )
\ea
where $x_{\alpha}=\e^{iS_{\alpha}/2h}$ and
$x_{2\beta-\alpha}=\e^{i(S_{\beta}-S_{\alpha}/2)/h+2i\delta}$. We define
$T=1/|t|^2=|g_1(-r)/g_1(r)|^2$ which is also equal to
\bb\label{TransR}\fl
T^{-1}=|<\bv(0)|\bv(0)+R^2\bv(0)+R^4\bv(0)+\cdots>|^2
=|<\bv(0)|(1-R^2)^{-1}\bv(0)>|^2
\ee
Only the even powers of $R$ contribute since to go trough $a'$ twice we need
to perform an even number of steps. Resumming the expression in~\eref{TransR}
involves the inverse of $(1-R^2)$ which can be computed from $(1-A^2)^{-1}$
since $R^2$ is simply the diagonal block matrix $\diag(A^2,A^2)$, and therefore
$(1-R^2)^{-1}=\diag((1-A^2)^{-1},(1-A^2)^{-1})$. After some algebra, we extract
the third component of $(1-R^2)^{-1}\bv(0)$ to obtain $T$
\bb
T=\left |
\frac{(1-x_{\alpha}x_{2\beta-\alpha})^2-q^2(x_{\alpha}\e^{-i\phi}-x_{
2\beta-\alpha}\e^{
i\phi})^2}{1-p^2x_{\alpha}x_{2\beta-\alpha}-q^2x_{\alpha}^2\e^{-2i\phi}}\right
|^2
\ee
There are two obvious cases. When $p=1$ and $q=0$, one obtains
$T=|1-x_{\alpha}x_{2\beta-\alpha}|^2=|1-\e^{iS_{\beta}/h+2i\delta}|^2$, or
$T=4\sin^2(S_{\beta}/2h+\delta)$, which was obtained previously in~\eref{Tbeta}.
Oppositely, when $p=0$ and $q=1$, the particle describes orbits around
$2\beta-\alpha$, and $T=|1-x_{2\beta-\alpha}^2\e^{2i\phi}|^2$, or
$T=4\sin^2[(S_{\beta}-\ff S_{\alpha})/h+\delta+\phi]$. This expression depends
on
$\phi$ explicitly.
%
\subsection{Simple solvable model for two-interacting orbits}
\begin{figure*}
\centering
\includegraphics[width=0.5\textwidth, clip,angle=0]{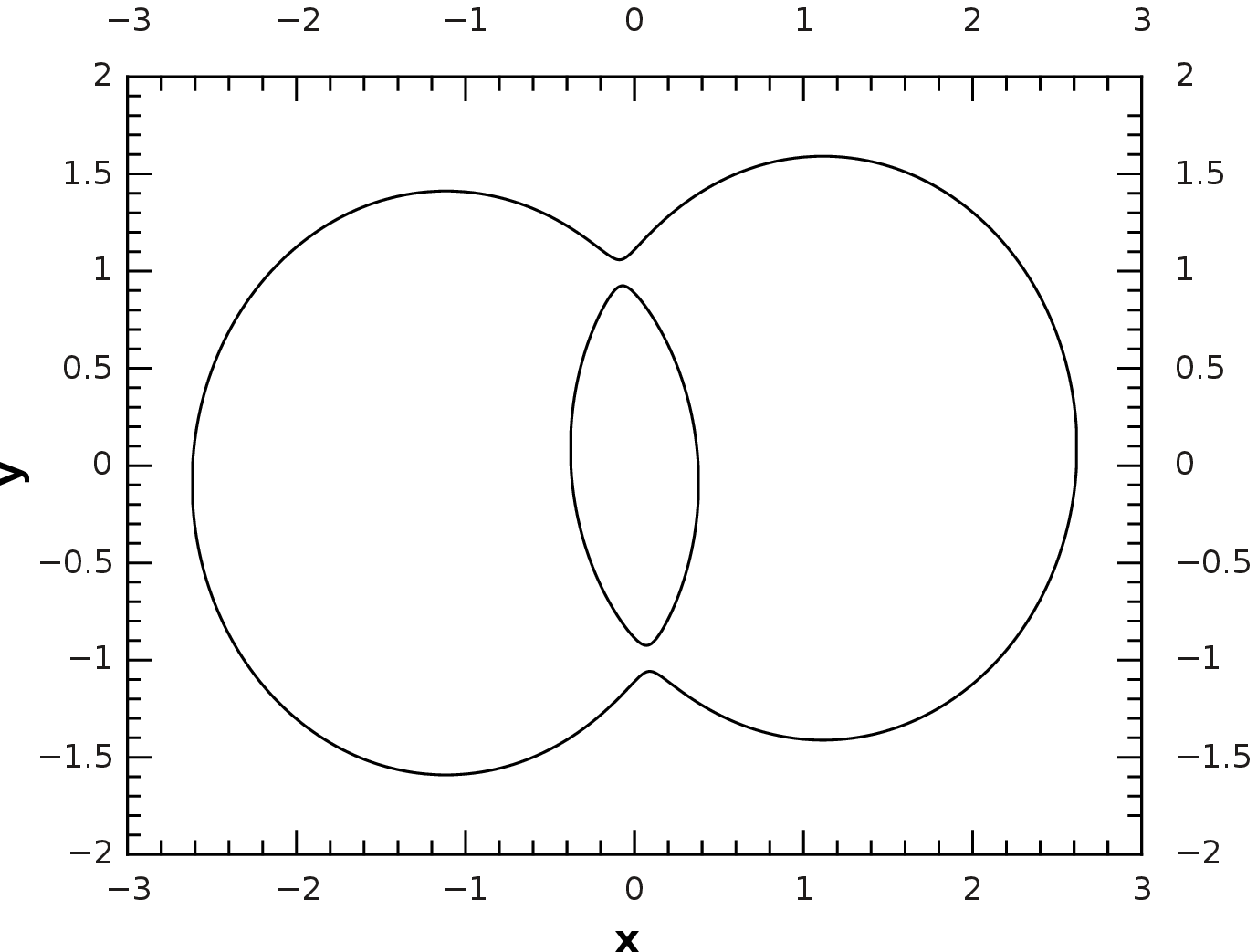}
(a)
\includegraphics[width=0.5\textwidth, clip,angle=0]{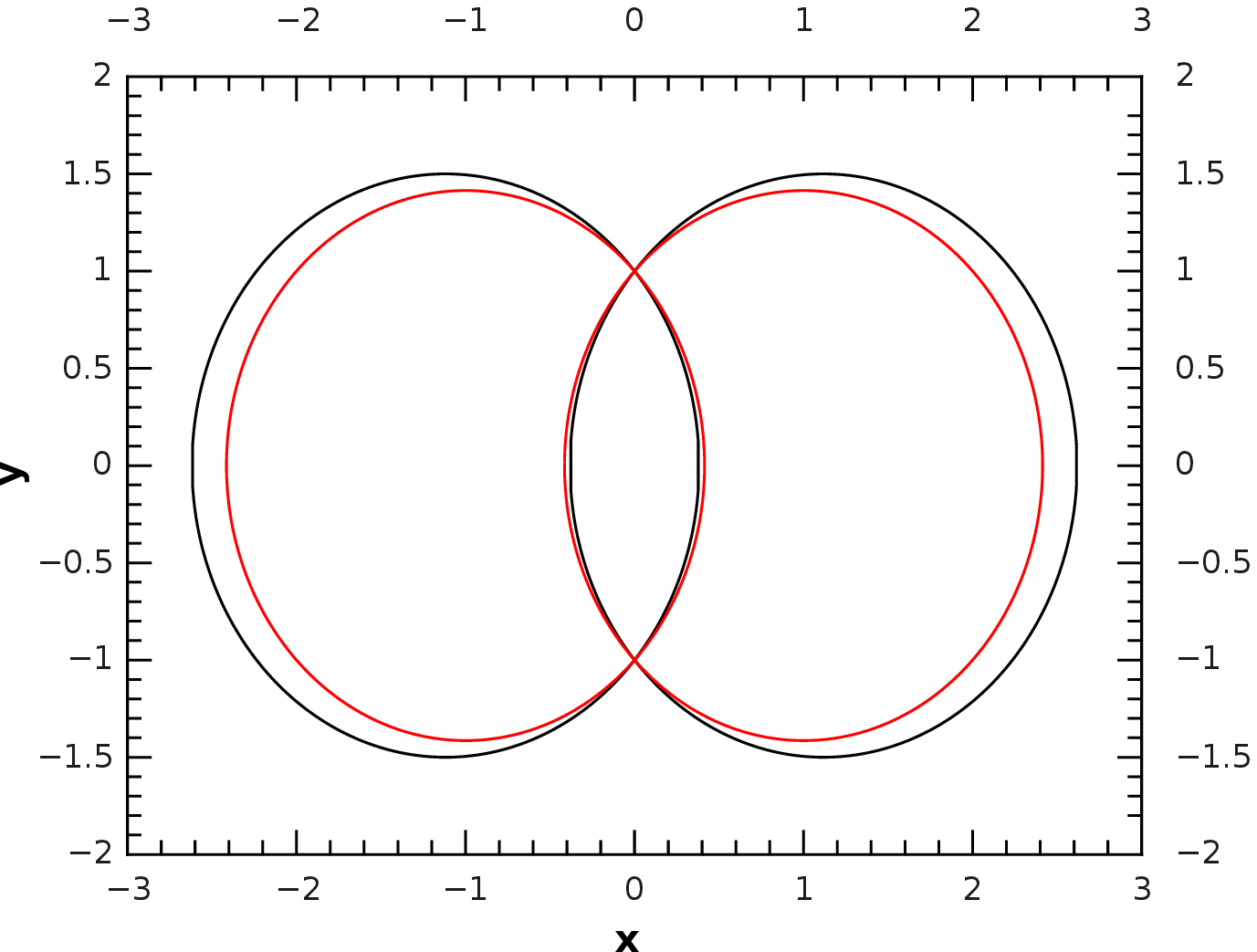}
(b)
\caption{(a) Fermi surface for $g=0.5$ and $\ds=0.1$, where a gap is present.
The two surfaces are tilted as their centers are not aligned on the horizontal
axis. (b)
Fermi surface for $g=0.5$ and $\ds=0$ (black), and $g=\ds=0$ (red). When $g\neq
0$, the area of the circular cyclotronic trajectories is slightly larger since
it is proportional to $r^2+g^2$.}
\label{fig7}
\end{figure*}

%
Let us rewrite the Hamiltonian~\eref{model_2} in the representation $(x,\hat
y=-ih\partial_x)$. One obtains the set of coupled differential equations
\bb
\nn
-h^2\partial^2_x\varphi_1+((x+1)^2-r^2)\varphi_1
+2\coupling\varphi_2=0,
\\ \label{eq_diff3}
2\coupling\varphi_1-h^2\partial^2_x\varphi_2+((x-1)^2-r^2)
\varphi_2=0
\ee
The advantage of this representation is that the imaginary parts
in~\eref{eq_diff2} are absent, at
the cost of a shift in
the harmonic potential. Function $\varphi_1$ is centered around $x=-1$ whereas
function $\varphi_2$ has dominant weight around $x=1$. We will consider
instead a slightly different set of equations
\bb
\nn
(\hat y+\ds)^2\varphi_1+((x+1)^2-r^2)\varphi_1
+2\coupling(x)\varphi_2=0,
\\ \label{eq_diff3bis}
2\bar\coupling(x)\varphi_1+(\hat y-\ds)^2\varphi_2+((x-1)^2-r^2)
\varphi_2=0
\ee
where $\ds$ is a parameter and the coupling $\coupling$ is a function of $x$:
$\coupling(x)=(x-i\ds)g$ with $g$ constant. The Hamiltonian operator $\hat H$
is then defined by
\ba\label{modelH2}
\hat H=\fff{1}{2}\left (
\begin{array}{cc}
(\hat y+\ds)^2+(x+1)^2-r^2 & 2g(x-i\ds)\\
2g(x+i\ds) & (\hat y-\ds)^2+(x-1)^2-r^2
\end{array} \right )
\ea
and the Fermi surface is the location of points given by the equation
\bb\fl\nn
H(x,y)=\fff{1}{4}[(y+\ds)^2+(x+1)^2-r^2][(y-\ds)^2+(x-1)^2-r^2]
\\
-g^2(x^2+\ds^2)=0
\ee
For $g$ and $\ds$ non zero, the surface is composed of two sheets separated by
a gap proportional to $\ds$, see~\efig{fig7}(a). It has to be noticed that for
this particular choice of coupling function, there is no observable gap on the
Fermi surface when $\ds=0$, since $\coupling(0)=0$, but the two surfaces
are still coupled at other points by $gx\neq 0$, see~\efig{fig7}(b). The
advantage of the Hamiltonian~\eref{modelH2} is that it can be factorized using
simple bosonic operators associated
with centers $\pm(1\pm i\ds)$ in the complex plane $(x,y)$:
\bb\nn
a=\frac{1}{\sqrt{2h}}\left (x+1+i\ds+h\partial_x \right ),\;
\ad=\frac{1}{\sqrt{2h}}\left (x+1-i\ds-h\partial_x \right ),
\\ \label{bosons}
b=\frac{1}{\sqrt{2h}}\left (x-1-i\ds+h\partial_x \right ),\;
\bd=\frac{1}{\sqrt{2h}}\left (x-1+i\ds-h\partial_x \right )
\ee
with $[a,\ad]=[b,\bd]=1$. The set of differential
equations~\eref{eq_diff3bis} are indeed identical to two coupled
harmonic oscillators
\bb\nn
h\left (\ad a+\ff \right
)\varphi_1+\coupling\varphi_2=\frac{r^2}{2}\varphi_1,
\\ \label{eq_diff3ter}
h\left (\bd b+\ff \right
)\varphi_2+\bar\coupling\varphi_1=\frac{r^2}{2}\varphi_2
\ee
and it is straightforward then to consider the following two-dimensional
'bosonic' operators
\ba\label{Poperators}
P=\left (
\begin{array}{cc}
a & \frac{g}{\sqrt{2h}} \\
\frac{g}{\sqrt{2h}} & b
\end{array} \right ),
P^{\dag}=\left (
\begin{array}{cc}
\ad & \frac{g}{\sqrt{2h}} \\
\frac{g}{\sqrt{2h}} & \bd
\end{array}\right )
\ea
to express the Hamiltonian as an extended harmonic oscillator in two-dimensions
\ba
\hat H
\left (
\begin{array}{c}
\varphi_1 \\
\varphi_2
\end{array} \right )=
\left \{
h P^{\dag}P+
\ff\left (
\begin{array}{cc}
h-r^2-g^2 & 0 \\
0 & h-r^2-g^2
\end{array} \right )
\right \}
\left (
\begin{array}{c}
\varphi_1 \\
\varphi_2
\end{array} \right )
=0
\ea
%
%
The 'bosonic' operators $P$ and $P^{\dag}$ satisfy the commutation relation
\ba\label{commP}
[P,P^{\dag}]=Q_0=
\left (
\begin{array}{cc}
1 & 2ig\ds/h \\
-2ig\ds/h & 1
\end{array} \right )=\sigma_0-2g\ds\sigma_2/h
\ea
which is not unity when the product $g$$\ds$ is not zero. We cannot therefore call
them 'bosonic' in the usual sense since there is a mixing of the two different
types of bosons due to the coupling. Here $\sigma_{i=0..3}$ are
the usual Dirac matrices in two dimensions~\footnote{We remind that the Dirac
matrices are defined by
$\sigma_0=\left( \begin{smallmatrix} 1 & 0\\ 0 & 1 \end{smallmatrix}
\right)$, $\sigma_1=\left( \begin{smallmatrix} 0 & 1\\ 1 & 0 \end{smallmatrix}
\right)$, $\sigma_2=\left( \begin{smallmatrix} 0 & -i\\ i & 0 \end{smallmatrix}
\right)$, and $\sigma_3=\left( \begin{smallmatrix} 1 & 0\\ 0 & -1
\end{smallmatrix}
\right)$}. There are two possible ways
to construct the wavefunctions, depending on the value of $\ds$. If $\ds=0$,
then $P$ and $P^{\dag}$ are true bosonic operators, and we can construct the
ground-state solution $P\Psi_0=0$ of lowest energy $E_0=\ff(h-r^2-g^2)=0$,
with $\Psi_0=(\varphi_1^{(0)},\varphi_2^{(0)})^\mathsf{T}/\sqrt{2}$. This
imposes the constraint $h=r^2+g^2$ on the field. Normally we construct the
states above the ground state energy by quantization of the area, or
$E_n=h(n+\ff)\propto r^2+g^2$, but here we keep $r$ constant (or constant Fermi
energy) and solve for $h$ values for which a set of bounded wavefunctions
can be found. It is easy to see that the first
component $\varphi_1^{(0)}$ satisfies the factorized differential equation
\bb
(x+h\partial_x\pm\sqrt{1+g^2})(x+h\partial_x\mp\sqrt{1+g^2})\varphi_1^{(0)}=0
\ee
The solutions are simple combinations of two Gaussian exponentials centered at
$\pm x_g=\pm \sqrt{1+g^2}$
\bb\fl
\varphi_1^{(0)}(x)=A\exp\left [-\frac{(x+x_g)^2}{2h} \right ]
+B\exp\left [-\frac{(x-x_g)^2}{2h} \right ],
\\ \nn\fl
\varphi_2^{(0)}(x)=-\frac{1-\sqrt{1+g^2}}{g}A\exp\left [-\frac{(x+x_g)^2}{2h}
\right ]
-\frac{1+\sqrt{1+g^2}}{g}B\exp\left [-\frac{(x-x_g)^2}{2h} \right ]
\ee
%
\begin{figure*}
\centering
\includegraphics[width=0.5\textwidth, clip,angle=0]{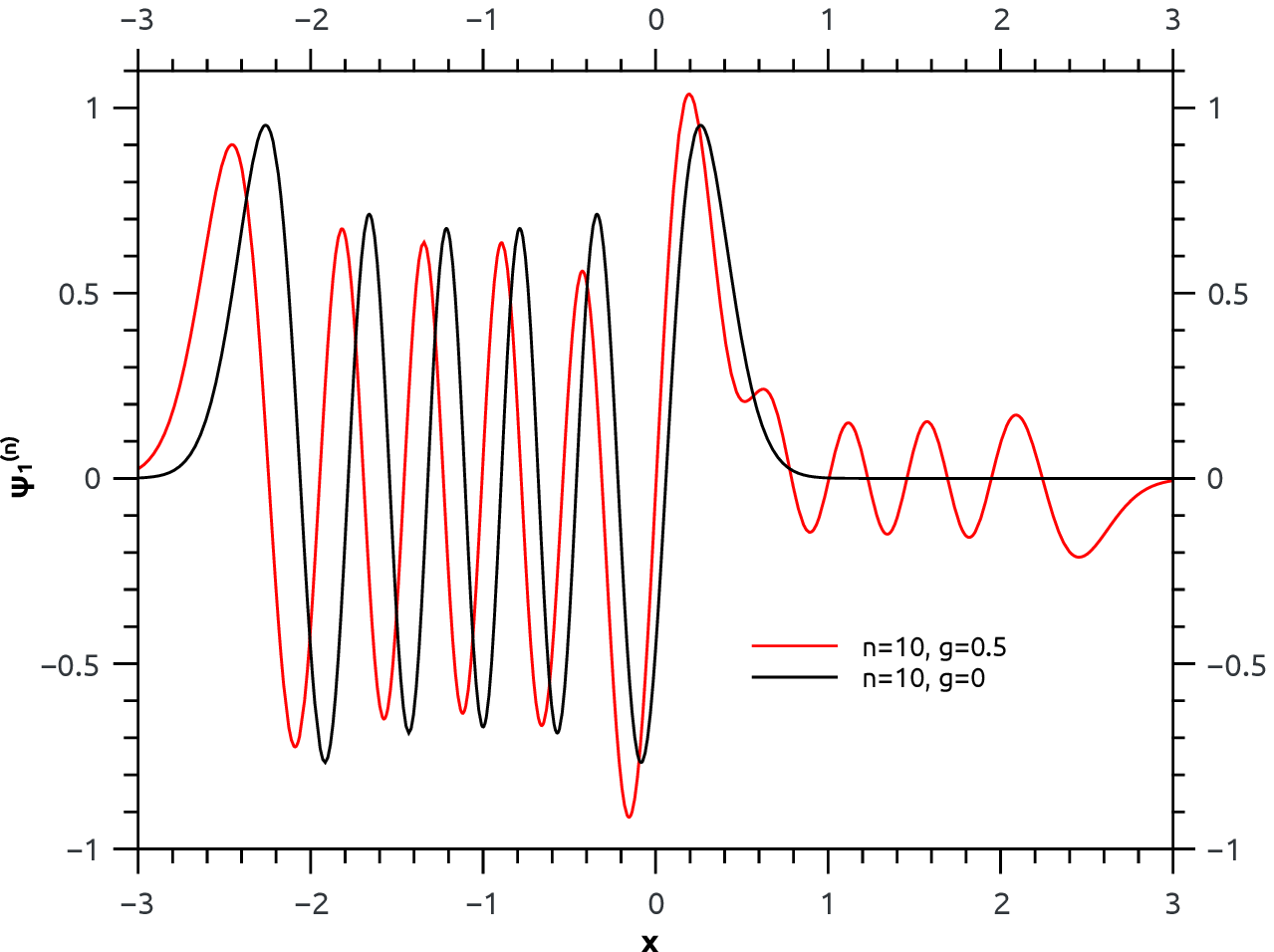}
\includegraphics[width=0.5\textwidth, clip,angle=0]{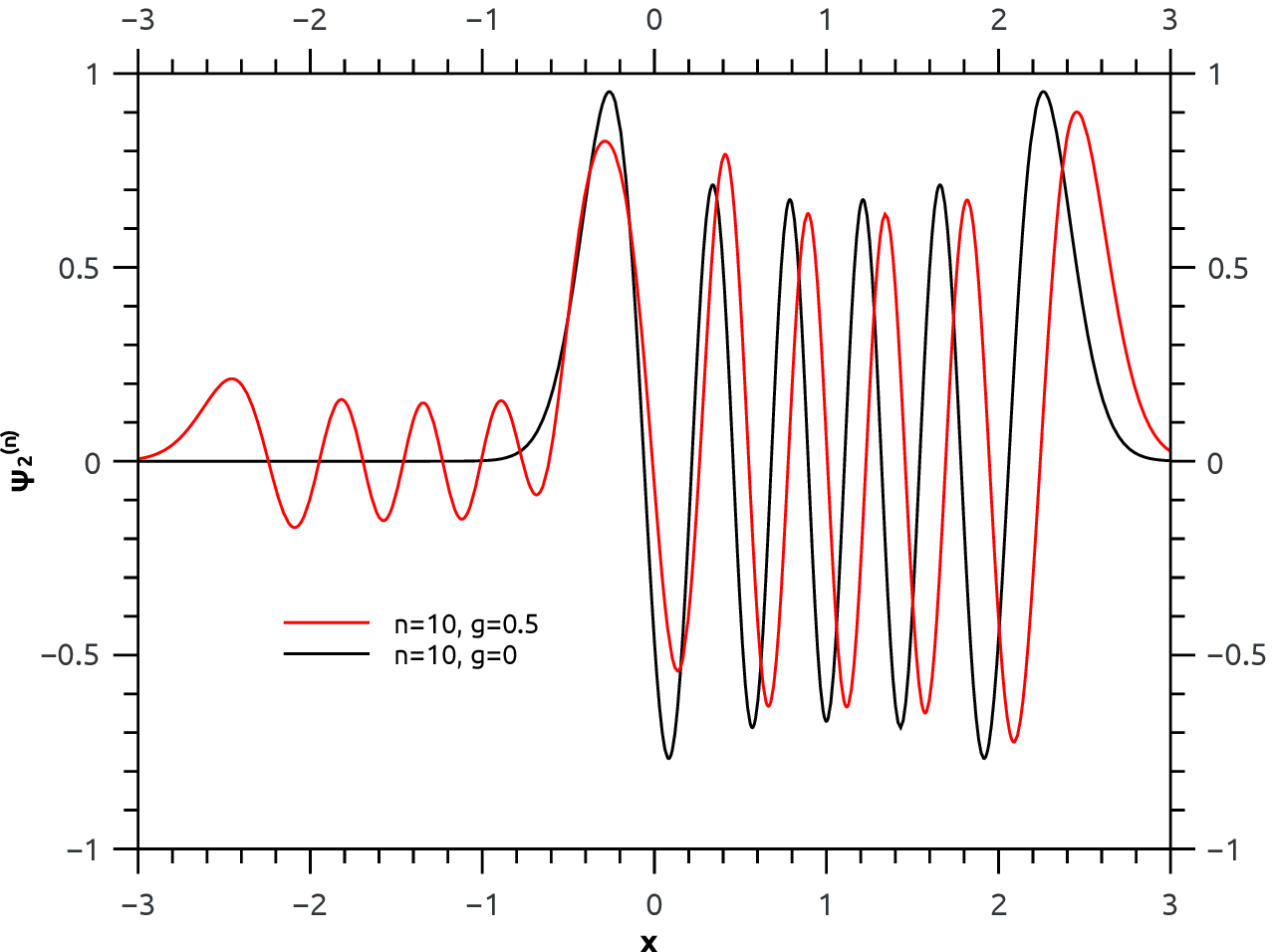}
\caption{Wave profile of bound states $\varphi_1^{(n)}$ and $\varphi_2^{(n)}$
for a coupling parameters $g=0.5$ and $\ds=0$ (red), at level $n=10$, and
comparison with
the free case ($g=0$ black, independent harmonic oscillators). For $g=0.5$ and
$g=0$, we take $h=(r^2+g^2)/(2n+1)$, corresponding to $h=0.107$ and
$h=0.095$ respectively. Constant $A=(\pi h)^{-1/4}$, and $B$
is deduced from~\eref{constantAB}.}
\label{fig8}
\end{figure*}
%
The two components are coupled together once the constants $A$ and $B$ are
determined. These constants satisfy a conservation equation,
depending on the filling factor. If we consider initially a system filled with
one electron in each orbital at zero coupling, therefore
two electrons in total, we impose that, by increasing the coupling, the
number of electrons per orbital does not change. One has the pair
of constraints $\int |\varphi_1^{(0)}|^2=\int |\varphi_2^{(0)}|^2=1$
(in this case we consider real functions), which leads to $<\Psi_0|\Psi_0>=1$,
and to the following relations of conservation
\bb\nn
\frac{1}{\sqrt{\pi h}}=A^2+B^2+2AB\e^{-x_g^2/h},
\\ \label{constantAB}
\frac{1}{\sqrt{\pi h}}=A^2\left (\frac{1-\sqrt{1+g^2}}{g}\right
)^2+B^2\left (\frac{1+\sqrt{1+g^2}}{g}\right
)^2-2AB\e^{-x_g^2/h}
\ee
The other state vectors at higher energy (or higher nodes) are given by the
successive application of $P^{\dag}$ on $\Psi_0$
\ba
\Psi_n=\frac{1}{\sqrt{2}}
\left (
\begin{array}{c}
\varphi_1^{(n)} \\
\varphi_2^{(n)}
\end{array} \right )
=\frac{1}{\sqrt{n!}}P^{\dag\,n}\Psi_0
\ea
with energy $E_n=h(n+\ff)-(r^2+g^2)/2$. When $E_n=0$, this imposes a field
value $h_n=(r^2+g^2)/(2n+1)$ for which $\Psi_n$ is solution
of~\eref{eq_diff3}.
In figure~\efig{fig8}, we have represented the two components $\varphi_1^{(n)}$
and $\varphi_2^{(n)}$ for the state $n=10$ at constant $r^2=2$. In the limit
of small coupling,~\eref{constantAB} leads to the solutions (we
choose $A>0$ and
$B<0$)
\bb
A\simeq (\pi h)^{-1/4},\;B\simeq -\frac{g}{2}(\pi h)^{-1/4}\rightarrow 0,
\\ \nn
\varphi_1^{(0)}\simeq (\pi h)^{-1/4}\exp\left [-\frac{(x+x_g)^2}{2h} \right ]
,\;
\varphi_2^{(0)}\simeq (\pi h)^{-1/4}\exp\left [-\frac{(x-x_g)^2}{2h} \right ]
\ee
which is expected for two independent orbitals. In general, the two constants
$A$ and $B$ are not independent because of~\eref{constantAB}, which leads to an
effective coupling between the two components of the wavefunction.

\begin{figure*}
\centering
\includegraphics[width=0.9\textwidth, clip,angle=0]{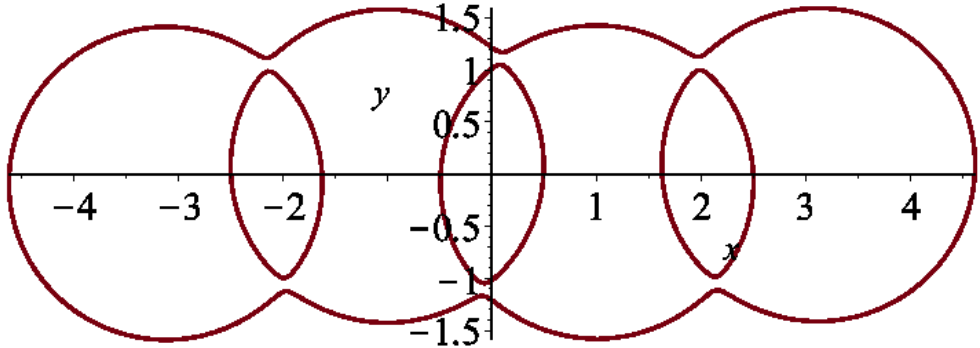}
\caption{Fermi surface of four individual coupled orbits, constructed from
operators~\eref{P4a} and~\eref{P4b}, with
coupling parameters $g=0.5$ and $\ds=0.1$.}
\label{fig9}
\end{figure*}

Let us now consider the case $\ds\neq 0$. The ground state is still defined by
$P\Psi_0=0$. Setting $z_g=\sqrt{(1+i\ds)^2+g^2}$, one obtains
\bb\fl
\varphi_1^{(0)}(x)=A\exp\left [-\frac{(x+z_g)^2}{2h} \right ]
+B\exp\left [-\frac{(x-z_g)^2}{2h} \right ],
\\ \nn\fl
\varphi_2^{(0)}(x)=-\frac{1+i\ds-z_g}{g}A\exp\left [-\frac{(x+z_g)^2}{2h}
\right ]
-\frac{1+i\ds+z_g}{g}B\exp\left [-\frac{(x-z_g)^2}{2h} \right ]
\ee
The conditions of normalization are the same as before, which leads to a
set of complex equations similar to~\eref{constantAB}.
The commutator~\eref{commP} prevents us to construct the excited
states $\Psi_n$, which satisfies $P^{\dag}P\Psi_n=n\Psi_n$, directly from
successive applications of $P^{\dag}$ on the ground state.
Instead we have to seek for linear combinations of functions $P^{\dag\,n}\Psi_0$
\bb\label{PsiR}
\Psi_n=R^{(n)}_nP^{\dag\,n}\Psi_0+R^{(n)}_{n-1}P^{\dag\,n-1}
\Psi_0+\cdots+R^{(n)}_0\Psi_0
\ee
where $R^{(n)}_k$ are constant matrices to be determined self-consistently. In
the limit $\ds\rightarrow 0$, only the matrix $R^{(n)}_n$ does not vanish, and
corresponds to the normalization factor. Computing $P^{\dag}P\Psi_n=n\Psi_n$
leads to a set of $(n+1)$ relations between these matrices at order $n$. In
particular,
by application of $P^{\dag}P$ on each element of~\eref{PsiR}, one has
\bb\nn
P^{\dag}PR^{(n)}_kP^{\dag\,k}\Psi_0=
\left (
[P^{\dag},[P,R^{(n)}_k]]-R^{(n)}_kQ_0\right )P^{\dag\,k}\Psi_0
+[P,R^{(n)}_k]P^{\dag\,k+1}\Psi_0
\\
+[P^{\dag},R^{(n)}_k]PP^{\dag\,k}\Psi_0+
R^{(n)}_kPP^{\dag\,k+1}\Psi_0
\ee
For the last two terms, after some algebra, we can move the operator $P$ to the
right of $P^{\dag\,k}$ and $P^{\dag\,k+1}$ using the binomial relation
\bb\nn
PP^{\dag\,k}\Psi_0=\sum_{l=0}^{k-1}\bin{k}{l}Q_{k-1-l}P^{\dag\,l}\Psi_0,\;
Q_l=[P^{\dag},Q_{l-1}],\;[P,P^{\dag}]=Q_0
\ee
The matrices $Q_k$ are zero when $Q_0=1$, and in this case we have simply
$PP^{\dag\,n}\Psi_0=nP^{\dag\,n-1}\Psi_0$.
The identification of each coefficient of $P^{\dag\,k}\Psi_0$ in the equation
$P^{\dag}P\Psi_n=n\Psi_n$ leads to the set of $(n+1)$ equations which are
composed of commutators. In particular, the first three equations read
\bb\nn
[P,R^{(n)}_n]=0,
\\ \label{solR}
[P,R^{(n)}_{n-1}]+nR^{(n)}_n(Q_0-1)=0,
\\ \nn \fl
[P,R^{(n)}_{n-2}]+R^{(n)}_{n-1}[(n-1)Q_0-n]
+[P^{\dag},[P,R^{(n)}_{n-1}]]
+n[P^{\dag},R^{(n)}_{n}]Q_0+\ff n(n+1)R^{(n)}_{n}Q_1=0
\ee
This can be solved for example using Dirac matrices with unknown scalar
coefficients. For
example, the matrix coefficients of the first excited state $n=1$,
$\Psi_1=(R_1P^{\dag}+R_0)\Psi_0$, can be found by solving the two equations
\bb\label{eqR01}
[P,R_1]=0,\;[P,R_0]=R_1(1-Q_0)
\ee
It is useful to write $P$ and $P^{\dag}$ using $2\times 2$ Dirac matrices
\bb\fl \nn
P=\frac{1}{\sqrt{2h}}\left [(x+h\partial_x)\sigma_0
+g\sigma_1+(1+i\ds)\sigma_3 \right ],\;
P^{\dag}=\frac{1}{\sqrt{2h}}\left [(x-h\partial_x)\sigma_0
+g\sigma_1+(1-i\ds)\sigma_3 \right ]
\ee
and separate the part proportional to identity from the remaining
$\sigma_i$'s: $P=(2h)^{-1/2}(x+h\partial_x)\sigma_0+P_0=D+P_0$ and
$P^{\dag}=(2h)^{-1/2}(x-h\partial_x)\sigma_0+P_0^{\dag}=D^{\dag}+P_0^{\dag}$,
with constant matrices
\bb\fl\label{P0}
P_0=\frac{1}{\sqrt{2h}}\left (
\begin{array}{cc}
1+i\ds & \coupling
\\
\coupling & -1-i\ds
\end{array} \right ),\;
P_0^{\dag}=\frac{1}{\sqrt{2h}}\left (
\begin{array}{cc}
1-i\ds & \coupling
\\
\coupling & -1+i\ds
\end{array} \right )
\ee
and $[D,D^{\dag}]=\sigma_0$. Differential operators $D$ and $D^{\dag}$ are
proportional to the identity matrix and commute with $P_0$ and $P_0^{\dag}$
which are constant matrices. Then the solutions of~\eref{eqR01} can be expressed
using $P_0$ and $P_0^{\dag}$ only. An obvious solution of the first equation is
$R_1=\alpha_0\sigma_0+\alpha_1 P_0$, where $\alpha_0$ and $\alpha_1$ are
constants which are determined by orthogonality and normalization of the
wavefunctions $\Psi_0$ and $\Psi_1$. Then a solution of the second
equation is simply $R_1=-(\alpha_0\sigma_0+\alpha_1 P_0)P_0^{\dag}$. In
particular, this leads to the factorization
\bb
\Psi_1=(\alpha_0\sigma_0+\alpha_1P_0)(P^{\dag}-P_0^{\dag})\Psi_0
=(\alpha_0\sigma_0+\alpha_1P_0)D^{\dag}\Psi_0
\ee
Writing the condition $<\Psi_0|\Psi_1>=0$ leads to
\bb
\alpha_0<\Psi_0|P_0^{\dag}\Psi_0>+\alpha_1<\Psi_0|P_0^{\dag}P_0\Psi_0>=0
\ee
The normalization $<\Psi_1|\Psi_1>=1$ gives a supplementary condition which
fixes the two constants (up to a phase factor)
\bb\fl
<P_0P_0^{\dag}>^2=|\alpha_0|^2\left (
<P_0^{\dag}P_0>^2(1+<P_0^{\dag}P_0>)
-(<P_0>^2+<P_0^{\dag}>^2)<P_0^{\dag}P_0>
\right .
\\ \fl \nn
\left .
-<P_0^{\dag}P_0>(<P_0><P_0P_0^{\dag}P_0>+<P_0^{\dag}><P_0^{\dag 2}P_0>)
-<P_0><P_0^{\dag}><(P_0^{\dag}P_0)^2>
\right )
\ee
where we have omitted $\Psi_0$ in the scalar products to simplify the
notations. When no coupling is present $\coupling=0$, $P_0=(1+i\ds)\sigma_3$,
and $P_0^{\dag}P_0=P_0P_0^{\dag}=(1+\ds^2)\sigma_0$. We also assume that in
this case that $<P_0>=<P_0^{\dag}>=0$, so that $|\alpha_0|^2=2$ and
$\alpha_1=0$, which corresponds to the uncoupled model of two electrons in two
independent orbits.
This method allows for the construction of all excited states and can
be generalized for a linear chain of $N$ coupled orbits. Indeed we can
represent the $P$ and $P^{\dag}$ operators as
extended matrix operators of dimension $N$ with coupling parameters $g$ and
$\ds$ similar to~\eref{Poperators}, and centers corresponding to each individual
oscillator. For example, in~\efig{fig9}, we have represented such surface, for
$N=4$ connected orbits, by considering the following extended bosonic operators
in
four dimensions
%
%
\ba\label{P4a}
P=\frac{1}{\sqrt{2h}}
\left (
\begin{array}{cccc}
x+3+i\ds+h\partial_x & g & 0 & 0
\\
g & x+1-i\ds+h\partial_x & g & 0
\\
0 & g & x-1+i\ds+h\partial_x & g
\\
0 & 0 & g & x-3-i\ds+h\partial_x
\end{array} \right )
\ea
and
\ba\label{P4b}
P^{\dag}=\frac{1}{\sqrt{2h}}
\left (
\begin{array}{cccc}
x+3-i\ds-h\partial_x & g & 0 & 0
\\
g & x+1+i\ds-h\partial_x & g & 0
\\
0 & g & x-1-i\ds-h\partial_x & g
\\
0 & 0 & g & x-3+i\ds-h\partial_x
\end{array} \right )
\ea
%
\section{Onsager phase of de Haas-van Alphen oscillations in linear chains of coupled orbits}
%
\begin{figure}                                                    
\centering
\resizebox{0.75\columnwidth}{!}{\includegraphics*{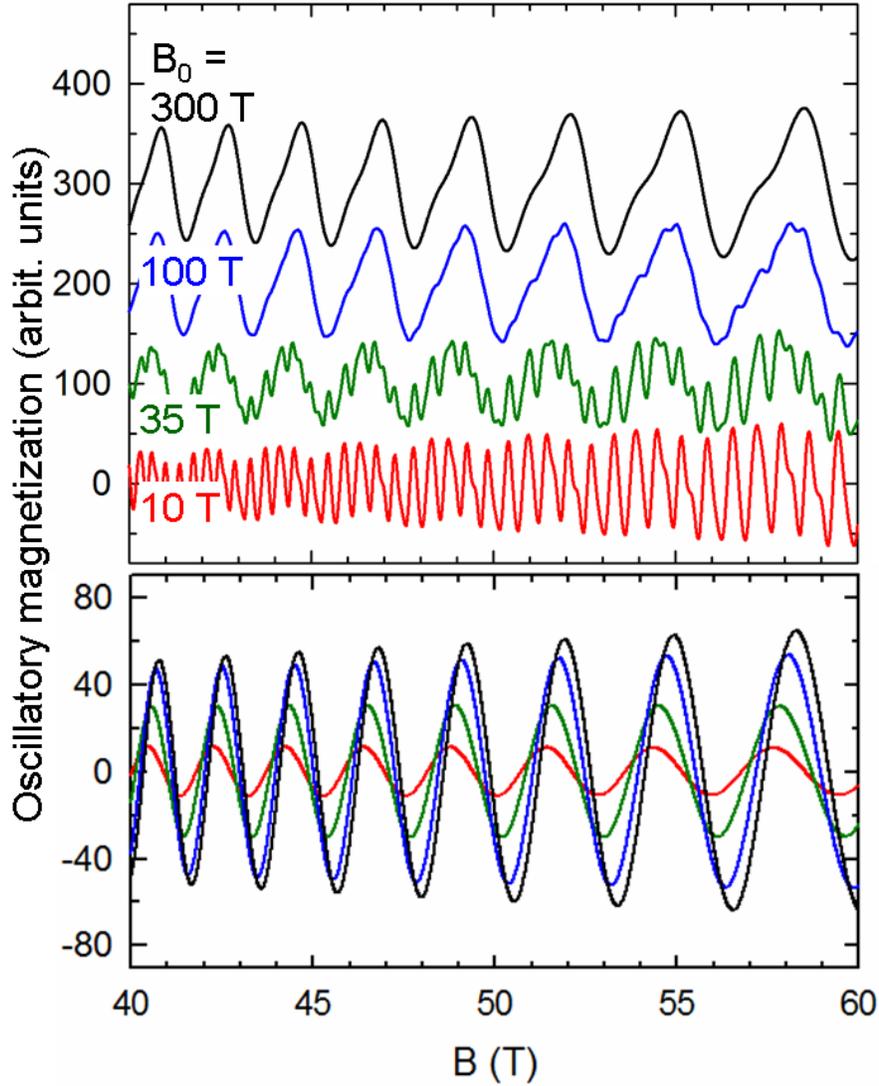}}
\caption{\label{fig:oscill} (a) De Haas-van Alphen oscillations calculated with
the parameters (effective masses, Dingle temperature, etc.) relevant to
$\theta$-(ET)$_4$CoBr$_4$(C$_6$H$_4$Cl$_2$)~\cite{Au12} albeit for various
values of the magnetic breakdown field $B_0$ ($B_0$ = 35 T holds for the
experimental data). Contribution of the component $\alpha$ is given in (b): as
$B_0$ increases, its amplitude increases and the Onsager phase shifts towards
high fields. }
\end{figure}
%
In this section, we consider de Haas-van Alphen oscillations observed in
quasi-two-dimensional organic metals with a Fermi surface which can be regarded
as a linear chain of orbits coupled by magnetic breakdown. Recall that Fourier
spectra of these compounds is composed of Fourier components, labeled $\eta$ in
the following, the frequency of which are linear combinations of that linked to
the closed orbit $\alpha$ and the magnetic breakdown orbit $\beta$: $F_{\eta}$ =
$n_{\alpha}F_{\alpha} + n_{\beta}F_{\beta}$. The field- and
temperature-dependent amplitude of several of these components does not follow
the usual Lifshitz-Kosevich formula due to oscillation of the chemical potential
in magnetic field. Nevertheless, Fourier amplitudes are accounted for by a
development up to the second order in damping factors in this case
~\cite{Au12,Au15,Au13}. An extensive discussion of this problematic is given in
Refs.~\cite{Au13b,Au14}. As an example, let us consider magnetic torque data
relevant to the organic
metal $\theta$-(ET)$_4$CoBr$_4$(C$_6$H$_4$Cl$_2$). Field- and
temperature-dependent de Haas-van Alphen oscillations amplitudes of this organic
metal are consistently accounted for by this formalism with the following
parameters: $F_{\alpha}$= 944 $\pm$ 4 T, $F_{\beta}$ = 4600 $\pm$ 10 T,
$m_{\alpha}$ = 1.81 $\pm$ 0.05, $m_{\beta}$ = 3.52 $\pm$ 0.19, $g^*_{\alpha}$ =
$g^*_{\beta}$ = 1.9 $\pm$ 0.2, $T_{D\alpha}$ = $T_{D\beta}$ = 0.79 $\pm$ 0.10 K,
$B_0$ = 35 $\pm$ 5 T, where  $F_{\alpha(\beta)}$, $m_{\alpha(\beta)}$,
$g^*_{\alpha(\beta)}$, $T_{D\alpha(\beta)}$ and $B_0$ are the frequencies,
effective masses, effective Land\'{e} factors, Dingle temperatures and magnetic
breakdown field, respectively~\cite{Au12}. Furthermore, the Onsager phase of
the various Fourier components is accounted for by Eq.~\ref{phi}, yielding
~\cite{Au13}
\begin{equation}
\label{Eq:phi}
\phi_{\eta} = \varphi_{\eta} - n^r_{\eta}\phi(B)
\end{equation}
where  $n^r_{\eta}$ is the number of reflections events and $\varphi_{\eta}$ is
equal to $\pi/2$ times the number of turning points of the $\eta$ orbit. De
Haas-van Alphen oscillations of Fig.~\ref{fig:oscill} are obtained with this set
of parameters, except that various values of $B_0$ are explored. As expected, as
$B_0$, hence the reflection probability $q$, increases, the amplitude of all the
components involving $\beta$ decreases and, at very high $B_0$, only remain the
contributions of $\alpha$ and its harmonics. The striking point, on which we
will focus in the following, is the observed shift of the $\alpha$ oscillations,
for which $n^r_{\alpha}=2$~\cite{Au13b,Au14}, as $B_0$ varies (whereas the
Onsager phase of $\beta$ oscillation remains unchanged since $n^r_{\beta}$ = 0
~\cite{Au13}).

Strictly speaking, the oscillations are not periodic in $1/B$ for finite $B_0$
values. This effect can be quantified considering an 'apparent frequency'
$F_{\textrm{app}}$ = $1/(B_i^{-1}-B_{i+1}^{-1})$ where the indexes $i$ and
$i+1$ mark
two successive oscillation maxima. According to Eq.~\ref{Eq:phi},
$F_{\textrm{app}}$ =$F_{\eta}$ + $(B_0/4\pi^2)d\phi/du$, yielding an
'universal' frequency shift:
\begin{equation}
\label{Eq:DeltaF}
\frac{\Delta F}{B_0} = \frac{1}{4\pi^2} \frac{d\phi_{\eta}}{du},
\end{equation}
where $\Delta F$ = $F_{\textrm{app}} - F$, which depends on $x$, e.g. on the
ratio
$B/B_0$, only, for a given $n^r_{\eta}$ value.
%
\begin{figure}
\centering
\resizebox{0.75\columnwidth}{!}{\includegraphics*{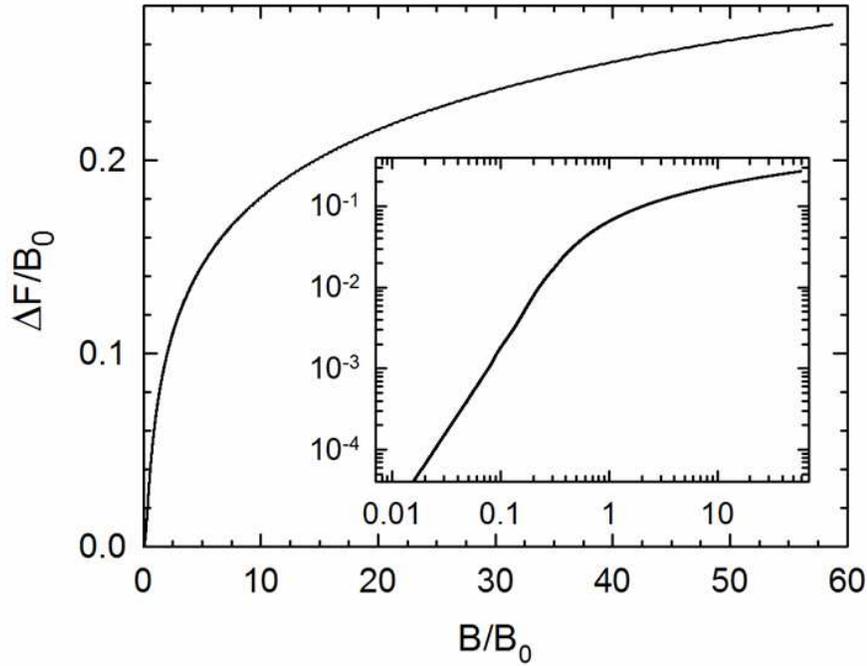}}
\caption{\label{fig:DeltaF} Field dependence of the 'apparent frequency' predicted by Eq.~\ref{Eq:DeltaF} for  $n^r_{\eta}$ = 2 which stands for $\alpha$ oscillations of the linear chain of coupled orbits. }
\end{figure}
%
Data of Fig.~\ref{fig:DeltaF} displays the frequency variations of the $\alpha$
component. Reported experimental data deal with magnetic fields of up to 56
T~\cite{Au13}, e.g. with maximum $B/B_0$ values of 1.6. According to the data of
Fig.~\ref{fig:DeltaF}, the corresponding frequency shift is $\Delta$F = 3 T
which is within the reported error bars (since $F_{\alpha}$= 944 $\pm$ 4 T for
the considered compound). Nevertheless, frequency shift predicted by
Eqs.~\ref{phi}, \ref{Eq:DeltaF} could be detected in the future at higher
magnetic fields and for orbits involving larger number of reflection events
$n^r_{\eta}$ such as observed in two-dimensional networks (see~\cite{La17}).
%
\section{Summary and Conclusion}
%
Calculation of transmission and reflection coefficients through a magnetic
breakdown junction have been reviewed with the aim of determining the Onsager
phase of de Haas-van Alphen oscillations. The problem of the phase divergence of
the S-matrix describing wave function transmission has been addressed by
suitable asymptotic analysis. Amplitude of the wave function was then
calculated, using approximate and exact models of connected Fermi surfaces,
yielding the field-dependent phase offset relevant to de Haas-van Alphen
oscillations for Fermi surfaces with magnetic breakdown. As a consequence,
experimental de Haas-van Alphen oscillations are not strictly periodic in
$B^{-1}$ for orbits with reflections at the magnetic breakdown junctions.
Nevertheless, frequency variations, which follow a 'universal' field dependence
remain small within realistic experimental conditions.
%
\section*{References}
%
\bibliographystyle{iopart-num}
\bibliography{tunneling}

\providecommand{\newblock}{}
\begin{thebibliography}{10}
\expandafter\ifx\csname url\endcsname\relax
  \def\url#1{{\tt #1}}\fi
\expandafter\ifx\csname urlprefix\endcsname\relax\def\urlprefix{URL }\fi
\providecommand{\eprint}[2][]{\url{#2}}

\bibitem{Mi99}
Mikitik G~P and Sharlai Y~V 1999 {\em Phys. Rev. Lett.\/} {\bf 82}(10)
  2147--2150

\bibitem{Fuchs2010}
Fuchs N~J, Pi{\'e}chon F, Goerbig O~M and Montambaux G 2010 {\em The European
  Physical Journal B\/} {\bf 77} 351--362

\bibitem{Wr13}
Wright A~R and McKenzie R~H 2013 {\em Phys. Rev. B\/} {\bf 87}(8) 085411

\bibitem{Fortin2015}
Fortin J~Y and Audouard A 2015 {\em The European Physical Journal B\/} {\bf 88}
  1--7

\bibitem{Slutskin:1967}
Slutskin A and Kadigrobov A 1967 {\em Sov. Phys. Solid State\/} {\bf 9}

\bibitem{Slutskin:1968}
Slutskin A 1968 {\em Sov. Phys. JETP\/} {\bf 26} 474--482

\bibitem{Huang:1976}
Huang W and Taylor P~L 1976 {\em Phys. Rev. Lett.\/} {\bf 36}(4) 231--233

\bibitem{Pi62}
Pippard A 1962 {\em Proc. R. Soc. London\/} {\bf A 270} 1

\bibitem{Os88}
Oshima K, Mori T, Inokuchi H, Urayama H, Yamochi H and Saito G 1988 {\em Phys.
  Rev. B\/} {\bf 38}(1) 938--941

\bibitem{Au13}
Audouard A, Fortin J~Y, Vignolles D, Lyubovskii R~B, Zhilyaeva E~I, Lyubovskaya
  R~N and Canadell E 2013 {\em Synthetic Metals\/} {\bf 171} 51 -- 55

\bibitem{Lyubovskii:2008}
Lyubovski{\u{\i}} R, Pesotski{\u{\i}} S, Biberacher W, Zhilyaeva E, Bogdanova A
  and Lyubovskaya R 2008 {\em Physics of the Solid State\/} {\bf 50} 1560--1564

\bibitem{Rosen:1932}
Rosen N and Zener C 1932 {\em Phys. Rev.\/} {\bf 40} 502--507

\bibitem{Chambers:1968}
Chambers W 1968 {\em Phys. Rev.\/} {\bf 165} 799--809

\bibitem{Torosov:2011}
Torosov B~T and Vitanov N~V 2011 {\em Phys. Rev. A\/} {\bf 84}(6) 063411

\bibitem{Au12}
Audouard A, Fortin J~Y, Vignolles D, Lyubovskii R~B, Drigo L, Duc F, Shilov
  G~V, Ballon G, Zhilyaeva E~I, Lyubovskaya R~N and Canadell E 2012 {\em EPL
  (Europhysics Letters)\/} {\bf 97} 57003--

\bibitem{Au15}
Audouard A, Fortin J~Y, Vignolles D, Lyubovskii R~B, Drigo L, Shilov G~V, Duc
  F, Zhilyaeva E~I, Lyubovskaya R~N and Canadell E 2015 {\em Journal of
  Physics: Condensed Matter\/} {\bf 27} 315601

\bibitem{Lam:1998}
Lam C 1998 {\em J.Math.Phys.\/} {\bf 39} 5543--5558 (\textit{Preprint}
  \eprint{hep-th/9804181})

\bibitem{Rojo:2010}
Rojo A~G Matrix exponential solution of the landau-zener problem equation (20)
  (\textit{Preprint} \eprint{http://arxiv.org/abs/1004.2914v1})

\bibitem{Kholodenko:2012}
Kholodenko A and Silagadze Z 2012 {\em Physics of Particles and Nuclei\/} {\bf
  43} 882--888

\bibitem{Holmes}
Holmes M~H 1995 {\em Introduction to perturbation methods\/} Texts in applied
  mathematics (New-York: Springer-Verlag) page 299

\bibitem{Abramowitz}
Abramowitz M and Stegun I 1984 {\em Pocketbook of mathematical functions\/}
  (Thun, Frankfurt am Main: Verlag Harri Deutsch) asymptotic forms for large
  argument in the Kummer function are given in 13.5.1. For small $h$, the two
  asymptotic expansions for $|y|\simeq r$ and $-r<y<r$ are given respectively
  by 13.5.19 and 13.5.21 with $\cos\theta=y/r$

\bibitem{Kochkin:1968}
Kochkin A 1968 {\em Sov. Phys. JETP\/} {\bf 27} 324--327

\bibitem{Hortacsu:2012}
Horta\c{c}su M 2012 {\em Heun functions and their uses in physics\/} (World
  Scientific Publishing Company) chap~2, pp 23--39 (\textit{Preprint}
  \eprint{arXiv:1101.0471})

\bibitem{Kaganov:1983}
Kaganov M and Slutskin A 1983 {\em Phys. Rep.\/} {\bf 98} 189--271 (See eq.
  5.22 and figures 17 and 19a)

\bibitem{Au13b}
Audouard A and Fortin J~Y 2013 {\em Comptes Rendus Physique\/} {\bf 14} 15 --
  26

\bibitem{Au14}
Audouard A and Fortin J~Y 2014 {\em Low Temperature Physics\/} {\bf 40}
  344--351

\bibitem{La17}
Laukhin V~N, Audouard A, Fortin J~Y, Vignolles D, Prokhorova T~G, Yagubskii E~B
  and Canadell E 2017 {\em Fiz. Nizk. Temp.\/} {\bf this issue}

\end{thebibliography}

\end{document}